\documentclass[a4paper,onecolumn, superscriptaddress,10pt, accepted=2021-11-27, issue=9, volume=5, shorttitle=papers/compositionality-5-9]{compositionalityarticle}
\pdfoutput=1

\usepackage[numbers,sort&compress]{natbib}
\usepackage{amsmath,amssymb,amsthm}
\usepackage{stmaryrd}
\usepackage{setspace}
\usepackage{tikz,pgf}
\usepackage{enumitem}
\usepackage{eufrak}
\usepackage{xurl}
\usetikzlibrary{shapes}
\usetikzlibrary{decorations.markings}
\usepackage{multirow,array}

\usepackage{tikzit}

\tikzstyle{circ}=[fill=white, draw=black, shape=circle]
\tikzstyle{blank}=[fill=white, draw=white, shape=circle]
\tikzstyle{none}=[fill=none, draw=none]
\tikzstyle{copy}=[fill=white, draw=black, shape=circle, minimum height=0.2cm, inner sep=0]
\tikzstyle{varCopy}=[fill=black, draw=black, shape=circle, minimum height=0.2cm, inner sep=0]
\tikzstyle{copy2}=[fill=black, draw=black, shape=circle, minimum height=0.2cm, inner sep=0]
\tikzstyle{1morph1}=[fill=white, draw=black, shape=rectangle, minimum width=1cm, minimum height=1cm]
\tikzstyle{1morph}=[fill=white, draw=black, shape=rectangle, minimum width=0.75cm, minimum height=0.75cm, inner sep=0.1cm]
\tikzstyle{2morph2}=[fill=white, draw=black, shape=rectangle, minimum width=1cm, minimum height=2cm]
\tikzstyle{2morph}=[fill=white, draw=black, shape=rectangle, minimum width=1cm, minimum height=1.25cm, inner sep=0.1cm]
\tikzstyle{nmorph}=[fill=white, draw=black, shape=rectangle, minimum height=6cm, minimum width=1cm, inner sep=0.1cm]
\tikzstyle{1state}=[fill=white, draw=black, regular polygon, regular polygon sides=3, minimum height=0.5cm, regular polygon rotate=-30]
\tikzstyle{dbox}=[fill=white, draw=black, dashed, shape=rectangle, minimum width=2cm, minimum height=1cm, inner sep=0.1cm]
\tikzstyle{vdbox}=[fill=white, draw=black, dashed, shape=rectangle, minimum width=2cm, minimum height=1.5cm, inner sep=0.1cm]
\tikzstyle{bigbox}=[fill=white, draw=black, dashed, shape=rectangle, minimum width=2cm, minimum height=4cm, inner sep=0.1cm]
\tikzstyle{2state}=[inner sep=0.05cm, fill=white, draw=black, isosceles triangle, minimum width=1.25cm, isosceles triangle apex angle=90, shape border rotate=180]
\tikzstyle{var2state}=[inner sep=0.05cm, fill=white, draw=black, isosceles triangle, minimum width=1.25cm, isosceles triangle apex angle=60, shape border rotate=180]
\tikzstyle{g2state}=[inner sep=0.05cm, fill=white, draw=black, isosceles triangle, minimum width=6cm, isosceles triangle apex angle=110, shape border rotate=180]
\tikzstyle{bigstate}=[inner sep=0.05cm, fill=white, draw=black, isosceles triangle, minimum width=3cm, isosceles triangle apex angle=110, shape border rotate=180]
\tikzstyle{bigeffect}=[inner sep=0.05cm, fill=white, draw=black, isosceles triangle, minimum width=3cm, isosceles triangle apex angle=110]
\tikzstyle{g2effect}=[inner sep=0.05cm, fill=white, draw=black, isosceles triangle, minimum width=6cm, isosceles triangle apex angle=110]
\tikzstyle{2effect}=[inner sep=0.05cm, fill=white, draw=black, isosceles triangle, minimum width=1.25cm, isosceles triangle apex angle=90]
\tikzstyle{b2effect}=[inner sep=0.05cm, fill=white, draw=black, isosceles triangle, minimum width=2cm, isosceles triangle apex angle=90]
\tikzstyle{midArrow}=[-, decoration={{markings,mark=at position .5 with {\arrow{>}}}}, postaction=decorate]

\tikzstyle{arrow}=[->]

\input{tikzit/tikzstyles.tikzdefs}

\let\phi\varphi
\let\epsilon\varepsilon
\let\theta\vartheta

\pgfdeclarelayer{nodelayer}
\pgfdeclarelayer{edgelayer}
\pgfdeclarelayer{foreground}
\pgfsetlayers{nodelayer,edgelayer,foreground}

\tikzstyle{circ}=[fill=white, draw=black, shape=circle]
\tikzstyle{blank}=[fill=white, draw=white, shape=circle]
\tikzstyle{none}=[fill=none, draw=none]

\tikzstyle{copy}=[fill=white,draw=black,shape=circle,minimum height =
0.2cm,inner sep=0]

\tikzstyle{varCopy}=[fill=black,draw=black,shape=circle,minimum height=0.2cm,inner sep=0]

\tikzstyle{copy2}=[fill=black,draw=black,shape=circle,minimum height =
0.2cm,inner sep=0]

\tikzstyle{1morph1}=[fill=white,draw=black,shape=rectangle,
minimum width=1cm,minimum height=1cm]

\tikzstyle{1morph}=[fill=white,draw=black,shape=rectangle,minimum
width=0.75cm,minimum height=0.75cm,inner sep=0.1cm]

\tikzstyle{2morph2}=[fill=white,draw=black,shape=rectangle,minimum
width = 1cm, minimum height = 2cm]

\tikzstyle{2morph}=[fill=white,draw=black,shape=rectangle,minimum
width=1cm,minimum height=1.25cm,inner sep=0.1cm]

\tikzstyle{nmorph}=[fill=white,draw=black,shape=rectangle,minimum height = 6cm, minimum
width = 1cm, inner sep=0.1cm]

\tikzstyle{1state}=[fill=white,draw=black,regular polygon, regular
polygon sides = 3, minimum height = 0.5cm, regular polygon rotate = -30]

\tikzstyle{dbox}=[fill=white,draw=black,dashed,shape=rectangle,minimum
width=2cm,minimum height=1cm,inner sep=0.1cm]

\tikzstyle{vdbox}=[fill=white,draw=black,dashed,shape=rectangle,minimum
width=2cm,minimum height=1.5cm,inner sep=0.1cm]

\tikzstyle{bigbox}=[fill=white,draw=black,dashed,shape=rectangle,minimum
width=2cm,minimum height=4cm,inner sep=0.1cm]

\tikzstyle{2state}=[inner sep=0.05cm,fill=white,draw=black,isosceles
triangle, minimum
width=1.25cm,isosceles triangle apex angle = 90,shape border rotate = 180]

\tikzstyle{var2state}=[inner sep=0.05cm,fill=white,draw=black,isosceles
triangle, minimum
width=1.25cm,isosceles triangle apex angle = 60,shape border rotate = 180]

\tikzstyle{g2state}=[inner sep=0.05cm,fill=white,draw=black,isosceles
triangle,minimum width = 6cm, isosceles triangle apex angle = 110,shape border rotate =180]

\tikzstyle{bigstate}=[inner sep=0.05cm,fill=white,draw=black,isosceles
triangle,minimum width = 3cm, isosceles triangle apex angle = 110,shape border rotate =180]

\tikzstyle{bigeffect}=[inner sep=0.05cm,fill=white,draw=black,isosceles
triangle,minimum width = 3cm, isosceles triangle apex angle = 110]

\tikzstyle{g2effect}=[inner sep=0.05cm,fill=white,draw=black,isosceles
triangle,minimum width = 6cm, isosceles triangle apex angle = 110]

\tikzstyle{2effect}=[inner sep=0.05cm,fill=white,draw=black,isosceles triangle,minimum
width=1.25cm, isosceles triangle apex angle = 90]

\tikzstyle{b2effect}=[inner sep=0.05cm,fill=white, draw=black, isosceles triangle,
minimum width =2cm, isosceles triangle apex angle=90]

\tikzstyle{arrow}=[->]

\tikzset{->-/.style={decoration={markings,mark=at position .5 with {\arrow{>}}},
postaction={decorate}}}

\newcommand{\A}{\mathcal{A}} 
\newcommand{\bAgent}{\mathcal{B}} 
\newcommand{\R}{\mathbb{R}} 
\newcommand{\eps}{\epsilon} 
\newcommand{\K}{\mathcal{K}} 
\renewcommand{\P}{\mathcal{P}} 
\newcommand{\pow}{\mathcal{P}} 

\newcommand{\br}{\mathrm{B}} 
\newcommand{\play}{\mathrm{P}} 
\newcommand{\rel}{\mathrm{Rel}} 
\newcommand{\G}{\mathcal{G}} 
\newcommand{\HH}{\mathcal{H}} 

\newcommand{\iset}{\{\star\}} 

\newcommand{\C}{\mathcal{C}} 
\newcommand{\D}{\mathcal{D}} 
\newcommand{\op}{\mathrm{op}} 
\newcommand{\set}{\mathbf{Set}} 
\newcommand{\id}{\mathrm{id}} 

\newcommand{\kl}{\mathbf{Kl}} 
\newcommand{\supp}{\mathrm{supp}} 
\newcommand{\update}{\mathcal{U}} 
\renewcommand{\exp}{\mathbb{E}} 

\renewcommand{\sc}{\mathcal{S}}

\newcommand{\conlens}{\mathbf{CL}} 
\newcommand{\congame}{\mathbf{ConGame}} 
\newcommand{\lens}{\mathbf{Lens}} 
\newcommand{\game}{\mathbf{Game}} 

\newcommand{\context}{\mathbb{C}} 

\newcommand{\leftc}{\mathcal{L}} 
\newcommand{\rightc}{\mathcal{R}} 

\newtheorem{theorem}[subsubsection]{Theorem}
\newtheorem{lemma}[subsubsection]{Lemma}
\newtheorem{proposition}[subsubsection]{Proposition}
\newtheorem{corollary}[subsubsection]{Corollary}

\theoremstyle{definition}
\newtheorem{example}[subsubsection]{Example}
\newtheorem{definition}[subsubsection]{Definition}
\newtheorem{notation}[subsubsection]{Notation}

\newcommand{\tikzStart}{\begin{tikzpicture}\begin{pgfonlayer}{foreground}}
\newcommand{\tikzEnd}{\end{pgfonlayer}{foreground} \end{tikzpicture}}
\newcommand{\fns}{\footnotesize}

\newcommand{\rb}{\raisebox}

\usepackage{hyperref}

\title{Bayesian open games}
\author{Joe Bolt}
\affiliation{}

\author{Jules Hedges}
\affiliation{Department of Computer and Information Sciences, University of Strathclyde, 26 Richmond Street, Glasgow, G11XH, U.K.}
\affiliation{20squares, \url{https://20squares.xyz}}
\affiliation{CyberCat Institute, \url{https://cybercat.institute}}
\email{jules.hedges@strath.ac.uk}

\author{Philipp Zahn}
\email{philipp@20squares.xyz}
\affiliation{20squares, \url{https://20squares.xyz}}
\affiliation{CyberCat Institute, \url{https://cybercat.institute}}
\date{}

\begin{document}

\maketitle

\begin{abstract}
This paper generalises the treatment of compositional game theory as introduced
 by Ghani et al.\ in 2018,  where  games are modelled as morphisms of a
symmetric monoidal category. From an economic modelling perspective, the notion
of a game in the work by Ghani et al.\  is not expressive enough for many
applications. This includes stochastic environments, stochastic choices by
players, as well as incomplete information regarding the game being played. The
current paper addresses these three issues all at once.
\end{abstract}

\section{Introduction}

In \cite{ghani2018compositional} the first compositional treatment of economic
game theory was introduced.
Following the literature on categorical open systems
\cite{fong_algebra_open_interconnected_systems}, \emph{open games} are modelled
as morphisms of a symmetric monoidal category.   

A distinctive and non-obvious feature of this approach is that the Nash
equilibrium condition \cite{nash50b}, one of the central concepts in classical
game theory to analyse rational behaviour of agents (cf. \cite[Chapter 1.2]{fudenberg1991game} and \cite[Chapter 2]{osborne1994course}), is itself compositional.

While an important first step, the treatment in \cite{ghani2018compositional} has two severe limitations:

\begin{enumerate}
\item Games are deterministic and as a consequence, there are no chance elements in
  the games and players have to choose deterministically.
\item Players have complete information about all relevant data of the
  game such as payoffs, number of players etc.
\end{enumerate}

Many interesting strategic situations feature chance elements.  Poker is one
example -- already discussed in the ground-breaking work of von Neumann 
and Morgenstern~\cite{neumann_theory_1944}. In an economic
context, the environment also often is non-deterministic. Two competing
companies face uncertain demand, exchange rates, lawsuits etc.

More subtle but also important is that players may need and may want to
randomise their actions. There are well known situations like Matching Pennies
(see, for instance, \cite[p.~16]{fudenberg1991game}) where playing
deterministically means being `beaten' all the time. Conceptually, from a game
theory perspective, this means that
there are games where equilibria do not exist when players are limited to
deterministic strategies (known as `pure strategies') whereas they do exist when players can choose stochastically (known as `mixed strategies').  

Lastly, it is a crude approximation to assume that players have
complete information. Examples abound. A used-car dealer knows how good the car
is that he is trying to sell to you. You may not know. Banks sitting on
toxic assets know how little value they actually have. The government trying to
buy these assets in order to save the financial system from collapse may not know. An
agent bidding in an auction may not know how many other bidders he competes
with. In most situations incomplete information is the norm and not the exception. 

The above limitations restrict the applicability of compositional game
theory to economic phenomena. And it restricts its usefulness for economists.
After all,  classical game theory already deals with these complications.

In this paper we provide a generalisation of open games which
solves the three problems above \emph{in one go}. We adapt the core definition of
compositional game theory such that the environment can be stochastic and
players can also choose in a non-deterministic fashion. Doing so, we also
introduce a way to deal with incomplete information.
Essentially we are lifting the same `trick', which has been introduced in
classical game theory to deal with games of incomplete information by John
Harsanyi \cite{harsanyi1967gamesI,harsanyi1968gamesII,harsanyi1968gamesIII}, to compositional game theory.

Harsanyi argued that instead of dealing with games of incomplete
information directly, which poses formidable conceptual problems, we can
transform such games into games of imperfect information by introducing the
notion of (game) types. Players have access to probability distributions
characterising these games as well as partial access to this information. For
instance, in an auction a player may know how much he values the good to be
auctioned. However, he may not know how other agents value the good. Assuming that
players update information according to Bayes' rule and adapting the
equilibrium notion of Nash, to what is called \emph{Bayesian Nash
equilibrium}, game theorists can work with such interactions no differently to
how they deal with chance elements in Poker. Thus by transforming the problem,
Harsanyi essentially opened the path to using tools that were more or less
already introduced by von Neumann and Morgenstern~\cite{neumann_theory_1944}. 

We are applying the same strategy. By introducing stochastic
environments and adapting the equilibrium notion from Nash to Bayesian Nash we show that our compositional
framework captures exactly Bayesian Games and thus allows to deal with
stochastic environments as well as with situations of incomplete information.
This distinguishes our work also from \cite{ghani2019compositional} which addresses
the issue of deterministic players in isolation.   

\subsection{Technical introduction}

Contrary at least to our own initial beliefs, addressing these issues requires some significant
adaptions of open games as defined in \cite{ghani2018compositional}.  

The recent understanding of open games has been based on \emph{lenses}, which consist of a pair of functions $X \to Y$ and $X \times R \to S$ packaged into a single morphism $(X, S) \to (Y, R)$ of a category.
Here, the function $X \to Y$ is the \emph{play function}, which plays out a given strategy by taking an initial state to a final state of the open game.
The function $X \times R \to S$, known as the \emph{coplay function} or \emph{coutility function}, is more subtle: It `backpropagates' payoffs  into the past, given an initial state.
This operation is `counterfactual', and the composition of lenses (which is not
trivial to define, nor is obvious to see is associative) intertwines ordinary forward and counterfactual (or `teleological') information flow.

An open game can then be viewed as a family of lenses indexed by a set of strategy profiles, together with another component describing which strategy profiles are Nash equilibria in a given \emph{context}.
A context for an open game consists of an initial state ($X$) and a function from final states to payoffs ($Y \to R$).
Contexts turn out also to be intimately connected to lenses, and indeed this was the initial hint that viewing open games in terms of lenses is a deep idea rather than a coincidence.

To someone trained in thinking about processes with side effects, it is entirely natural to begin by inserting a (finite support) probability monad $D$, and take the components of the lenses to be Kleisli morphisms $X \to D (Y)$ and $X \times R \to D (S)$, or equivalently to use lenses over the category of sets and (finite support) stochastic functions.
This allows the strategies of an open game to describe probabilistic behaviours, which are known as \emph{behavioural strategies} in game theory.
Unfortunately this doesn't work: In order to prove that lenses form a 
category (i.e.~are associative and unital) it is necessary that the forwards maps $X \to Y$ are homomorphisms of copying, and in the category of stochastic processes this characterises those processes that are actually deterministic.

Fortunately this problem has already been solved in the theory of lenses, although the solution is far from obvious.
We use the \emph{existential lenses} or \emph{coend lenses} as developed by Riley \cite{mitchell}.
This means we replace the pair or functions $X \to D (Y)$ and $X \times R \to D (S)$ with three things: A choice of set $A$, a function $X \to D (A \times Y)$ and a function $A \times R \to D (S)$.
Moreover a certain equivalence relation needs to be imposed, and this is precisely given by the following \emph{coend} \cite{loregian} (one of the universal constructions of category theory):
\[ \int^{A} (X \to D (A \times Y)) \times (A \times R \to D (S)) . \]
The proofs in this paper make heavy use of a diagrammatic language for existential lenses developed in \cite{mitchell}.

The second question is what should be considered a context of a Bayesian 
open game, i.e.~a replacement for the pair $X \times (Y \to R)$.
There is an existing characterisation of these contexts in terms of deterministic lenses, namely as a `state' lens $(1, 1) \to (X, S)$ and a `costate' lens $(Y, R) \to (1, 1)$.
However this turns out to be a red herring: in Section~\ref{sec:genGames} we show that generalising from this causes the category of open games to fail to be monoidal in an unexpected way.

It turns out that the appropriate notion of context consists of three 
things: a set $\Theta$ of \emph{unobservable} states, a joint distribution on $\Theta \times X$ (i.e.~an element of $D (\Theta \times X)$) and a function $\Theta \times Y \to D (R)$.
Again we need to impose a certain equivalence relation, which again turns out to be precisely a coend.
Remarkably this is equivalent to a state in the category of \emph{double 
lenses}, i.e.~lenses over the category of lenses.
This brings an unexpected theoretical unity to Bayesian open games, and means that the graphical language of \cite{mitchell} can be used throughout.

\section{Concrete open games}\label{chap:coGames}

We begin with a self-contained introduction to deterministic open games.
In essence, we will introduce the necessary machinery so that we can represent
simple classical games with diagrams as depicted below.

\begin{center}
	\tikzStart\input{tikz/openSequential}\tikzEnd
\end{center}

This diagram displays an interaction between two agents, A and B. Player
A moves first; player B observes the choice by A and then moves afterwards.
Both moves are consumed by an environment $c_k$ which provides the payoff for both
players. 

To get to a full understanding of this diagram (and deterministic open games),
in this section  we introduce several building blocks. Roughly, they can be
classified in two kinds.

First, we need a way to architecture the
information flow. As we will see in Sections \ref{sec:lens} to 
\ref{sec:conMonLens}, \emph{lenses} play a
crucial role by providing us with a categorical structure on which to build open
games.

Second, we need to flesh out the internals of the boxes in the diagram. Specifically, how does strategic
reasoning actually take place?  Central here is the notion of an \emph{agent}
who makes observations and chooses
a course of action. As we will see in Section
\ref{sec:conAgents} and Section \ref{brLens}, the key insight is to model an
agent as choosing against a \emph{context} which comprises how the environment
reacts to an agent's choices. The context is also the glue that keeps the
outside information flow and the internal reasoning together.  

Once we have introduced all the relevant parts, we will come back to the example
above. 

Note: The exposition in this section slightly deviates from previous work. We believe this eases the way
for the generalisations to come in Section \ref{chap:ogames} and thereafter.

\subsection{Lenses}\label{sec:lens}

The history of mathematical lenses is complicated, involving many
independent discoveries and fresh starts across numerous areas of
mathematics and computer science \cite{avigad98,depaiva_dialectica_categories_report,oles_category_theoretic_approach_semantics_programming_languages,hoffman_pierce_positive_subtyping,foster_etal_combinators_bidirectional_tree_transformations,palmer_making_haskell_nicer_game_programming}. An in-depth description of this
history can be found at \cite{lens}. We use lenses to describe the flow of information
through a game. A
lens for a given game describes which players have access to what
information when making a strategic decision, and also how information about players'
strategic decisions is ultimately fed into the outcome function for the game.  For example, it may
specify an order of play, or whether two players are playing in
parallel, or even whether some players are privy to certain
information in the environment that other players are not.

In general, lenses can be thought of as processes that perform some
computation and then propagate some resulting feedback from the environment backwards
through a system of which they are a part. In particular, this means
that lenses have both covariant and contravariant components. The
covariant component carries out the initial computation and the
contravariant component propagates the resulting feedback back through
the system.
Crucially, lenses are also \emph{compositional} in the sense that they admit both
sequential and parallel composition and, consequently, form a symmetric monoidal category. 

The lenses used in this paper are direct descendants of the lenses of
database theory.
Given some data $x$ of type $X$ we may want to view some part of it
$y$ of type $Y$. This is encapsulated by a \textit{view function} $v:
X\rightarrow Y$. From this `close-up' view of the database we may want
to edit the database by updating $y$. Given an update of the view $y$ we then need to know
how  this update propagates to an
update of the original data $x$. That is, given initial data
$x$ and an updated view $y': Y$, we should specify some updated $x':X$
given by some \textit{update function} $u : X\times Y \rightarrow X$.
The pair
$(v,u)$ is a \textit{lens} with type $X\rightarrow Y$. The connection
to our previous abstract definition of lenses is as follows:
\begin{itemize}
\item The covariant computation associated with the lens is the view function
$v: X\rightarrow Y$,
\item the resulting feedback from the environment is the update made to the
subdatabase returned by the view function, and
\item this feedback is propagated back to the whole database via the
  update function $u : X\times Y \rightarrow X$.
\end{itemize}
Abstracting away from databases, there is no reason to demand that the
feedback generated by the environment will have the same type as the
output of the lens computation.
Similarly, we may be interested in cases where the update function is
not-so-literally an `update' function, but merely a function that
propagates \emph{some kind} of feedback back through the system. As such, the
lenses we will be using will have types of the form $(X,S)\rightarrow
(Y,R)$ where the covariant component of the lens is of type
$X\rightarrow Y$ and the contravariant component is of type $X\times
R\rightarrow S$.

In game theory, we can regard players as `lenses that care about the
feedback they receive from the environment'. In a game with sequential
play, players make some play (computation), receive some utility
(feedback) from the outcome function, and then pass some feedback to
earlier players in the game (their outcome function given the moves that the
later players chose). Moreover, given that lenses admit parallel
composition as well as sequential composition, we obtain a
nuanced notion of information flow in a game. 

In the next subsections we describe a symmetric monoidal category of
concrete lenses. `Concrete' here refers to the fact that the view and
update functions are functions in $\set$. We then come to the core of this
section, the definition of a \emph{concrete open game}.  

\subsection{The category of concrete lenses}\label{sec:conLensCat}

\begin{definition}[Concrete lens]
Let $X,S,Y$ and $R$ be sets. A \textit{concrete lens} $l:
(X,S)\rightarrow (Y,R)$ is a pair of functions $(l_v : X\rightarrow Y,\:
l_u : X\times R \rightarrow S)$.
\end{definition}

As a trivial first example, there is an obvious mapping that takes a morphism of $\set\times\set^\op$ and returns a
concrete lens.

\begin{example}
	Let $f : X\rightarrow Y$ and $g: R\rightarrow S$. Define a concrete lens $\langle
	f,g\rangle : (X,S)\rightarrow (Y,R)$ by
	\begin{gather*}
		\langle f,g\rangle_v = f
		\\
		\langle f,g\rangle_u (x,r) = g(r)  . 
	\end{gather*}
\end{example}

\begin{definition}[Sequential composition of concrete lenses]\label{seqclens}
Let $l : (X,S)\rightarrow (Y,R)$ and $t: (Y,R) \rightarrow (Z,Q)$ be
concrete lenses. The \textit{sequential composite} $t\circ l :
(X,S)\rightarrow (Z,Q)$ is given by $\big( (t\circ l)_v : X\rightarrow
Z, (t\circ l)_u : X\times Q \rightarrow S \big)$ where 
\[
(t\circ l )_v = t_v \circ l_v
\]
and $(t\circ l )_u$ is given by
\begin{center}\tikzStart
	\node (a) {$X\times Q$};
	\path (a.east) -- +(2,0) node (b) {$X\times X\times Q$};
	\path (b.east) -- +(2,0) node (c) {$X\times Y\times Q$};
	\path (c.east) -- +(2,0) node (d) {$X\times R$};
	\path (d.east) -- +(2,0) node (e) {$S.$};

	\draw [->] (a.east) --  node [label=above:{\footnotesize $\Delta_X \times
	\id_X$}] {} (b.west);
	\draw [->] (b.east) -- node [label=above:{\footnotesize$\id_X\times l_v
	\times \id_Q$}] {} (c.west);
	\draw [->] (c.east) --  node [label=above:{\footnotesize$\id_X \times
	t_u$}] {} (d.west);
	\draw [->] (d.east) -- node [label=above:{\footnotesize$l_u$}] {} (e.west);
\tikzEnd\end{center}
As a string diagram $(t\circ l)_u$ is given by
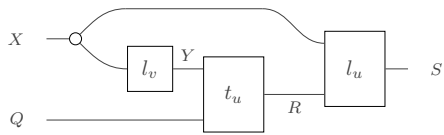
\begin{center}
\tikzStart
\scalebox{0.8}{\input{tikz/seqconlens.tex}}
\tikzEnd
\end{center}

\end{definition}

\begin{lemma}[Sequential composition of concrete lenses is
	associative]\label{cLensAssoc}
Suppose we have concrete lenses
\begin{center}\tikzStart
\node (l1) {$(X,S)$};

\path (l1.east) -- +(1,0) node (l2)[anchor=west] {$(Y,R)$};
\path (l2.east) -- +(1,0) node (l3)[anchor=west] {$(Z,Q)$};
\path (l3.east) -- +(1,0) node (l4)[anchor=west] {$(W,T)$};

\draw[->] (l1.east) -- node[above]{\fns$l$} (l2.west);
\draw[->] (l2.east) -- node[above]{\fns$m$} (l3.west);
\draw[->] (l3.east) -- node[above]{\fns$n$} (l4.west);
 
\tikzEnd  \end{center}
Then $n\circ (m\circ l) = (n\circ m )\circ l$.
\end{lemma}

\begin{theorem}[Concrete lenses form a category]
	There is a category $\conlens$ with pairs of sets as objects
	and concrete
	lenses as morphisms.
\end{theorem}

\subsection{The monoidal structure of concrete lenses}\label{sec:conMonLens}

\begin{definition}[Tensor composition of concrete lenses]
Let $l_1 : (X_1,S_1)\rightarrow (Y_1, R_1)$ and $l_2 :
(X_2,S_2)\rightarrow (Y_2,R_2)$ be concrete lenses. The \textit{tensor
composition} $l_1 \otimes l_2 : (X_1\times X_2, S_1\times S_2)
\rightarrow (Y_1 \times Y_2, R_1\times R_2)$ is given by $\big(
(l_1\otimes l_2)_v, (l_1 \otimes l_2)_u \big)$ where
\[
(l_1 \otimes l_2 )_v = l_{1_v} \times l_{2_v}
\]
and  $(l_1\otimes l_2)_u$ is given by
\bigskip
\begin{center}
\tikzStart
\node (a) {$X_1\times X_2 \times R_1 \times R_2$};
\path (a.east) -- +(3,0) node (b) {$X_1\times R_1\times X_2\times R_2$};
\path (b.east) -- +(2,0) node (c) {$S_1\times S_2 .$};

\draw [->] (a.east) -- node[label=above:{\footnotesize $\cong$}] {} (b.west);
\draw [->] (b.east) -- node[label=above:{\footnotesize $l_{1_u} \times
l_{2_u}$}] {} (c.west);
\tikzEnd
\end{center}
In a diagram, $(l_1\otimes l_2)_u$ is
\begin{center}\tikzStart
	\input{tikz/idk}
\tikzEnd\end{center}
\end{definition}

\begin{lemma}
$\otimes$ is a functor.
\end{lemma}

\begin{theorem}
There is a symmetric monoidal category $\conlens$ where the objects
are pairs of sets and the morphisms are concrete lenses. Sequential
composition and the monoidal tensor are as in the above definitions.
The monoidal unit is $I= (\{*\},\{*\})$.
\end{theorem}

The following observations about states and effects in $\conlens$ will be useful in the
remainder of this section.

\begin{lemma}\label{conStates}
	$\conlens \big( I, (X,S)\big) \cong X$.
\end{lemma}
\begin{proof}
	This is easily seen, as a state $l \in \conlens\big(I,(X,S)\big)$ is given by a
	pair 
	\[\big(
		s :
	\iset\rightarrow X, e : \iset\times S\rightarrow \iset\big).\]
\end{proof}

\begin{lemma}\label{conEffects}
	$\conlens \big( (Y,R), I \big) \cong (Y\rightarrow R)$
\end{lemma}
\begin{proof}
	An effect $l \in \conlens\big( (Y,R), I\big)$ is given by a pair
	\[
		\big(v : Y\rightarrow\iset, u : Y\times\iset\rightarrow R\big).
	\]
\end{proof}

\subsection{Concrete open games}\label{sec:conGames}

Now we have the necessary prerequisites in place to introduce the notion of a \textit{concrete open game}. A concrete open game
consists of a set of strategy profiles; a family of concrete lenses indexed by
the set of strategy profiles; and a best-response function. 

\begin{definition}[Concrete open game]
	Let $X,S,Y,$ and $R$ be sets.	A \textit{concrete open game} $\G :
	(X,S)\rightarrow (Y,R)$ is given by
	\begin{enumerate}
		\item A set of \textit{strategy profiles} $\Sigma$;
		\item A \textit{play function} $\play : \Sigma\rightarrow
			\conlens\big((X,S),(Y,R)\big)$; and
		\item A \textit{best-response function} $\br : X\times (Y\rightarrow
			R)\rightarrow \rel(\Sigma)$.
	\end{enumerate}
\end{definition}

Here $\rel (\Sigma) = \mathcal P (\Sigma \times \Sigma)$ is the set of binary relations on $\Sigma$. In this paper we will usually specify a relation in terms of its forward images $\Sigma \to \mathcal P (\Sigma)$.

The type $X$ is the type of \textit{observations} made by the game; the type $Y$ is the
type of \textit{actions} that can be chosen; the type $R$ is the type of
\textit{outcomes}; and the type $S$ is the type of \textit{co-outcomes}. Of the four types
associated with a concrete open game, the type $S$ is the most mysterious. Succinctly, its purpose is
to relay information about outcomes to games acting earlier. In a sequential composite
$\HH\circ\G$ of open games (we will define sequential composition of concrete open games
shortly), the
co-outcome type of $\HH$ is also the outcome type of $\G$. We think of $\HH$ as receiving some
outcome which is then acted upon by the contravariant component of a concrete lens given
by $\HH$'s play function before being passed back to $\G$ as $\G$'s outcome. 

The best-response function of an open game is an abstraction from the utility functions of classical game theory. Recall that a
Nash equilibrium for a normal-form game is a strategy profile in which no player has
incentive to unilaterally deviate. We can instead think of a relation on the set of
strategy profiles for a normal-form game where strategy profiles $\sigma$ and $\tau$ are
related if $\tau$ is the result of players unilaterally deviating from $\sigma$ to their
most profitable unilateral deviation. Nash equilibria are then the fixed points of this
relation. For convenience, in the definition of a concrete open game we work directly with a best-response
relation rather than preference relations.

The play function takes a strategy as argument and returns a concrete lens that
describes an \textit{open play} of the game $\G$ (`open' here means `lacking a particular
observation and outcome function' and is explained in the next paragraph). To justify this interpretation, recall
that a concrete lens $l : (X,S) \rightarrow (Y,R)$ consists of $v : X\rightarrow Y$ and $u
: X\times R\rightarrow S$. The view function $v$ describes how a game decides on an action
given an observation (similar to how strategies for sequential games work). The update
function $u$ describes precisely how games relay information about outcomes to other games acting
earlier.

As the name suggests, concrete open games are \emph{open} to their environment. The
appropriate notion of a \textit{context} for a concrete open game is given in the
following definition. A concrete open game together with a context can be thought of as a
full description of a game.
\begin{definition}
	Let $\G : (X,S)\rightarrow (Y,R)$. A \textit{history} for $\G$ is an element $x$
	of $X$, an \textit{outcome function} for $\G$ is a function $k : Y\rightarrow R$,
	and a \textit{context} for $\G$ is a pair $(x,k) : X\times (Y\rightarrow R)$.
\end{definition}

We are now in a position to justify the type of the best-response function. The best
response functions takes a context as argument, and a context is precisely the information
required for resolving the `openness' of a concrete open game. Given a context, the best
response function then returns the set of best deviations from a strategy profile
$\sigma$.  

We represent a concrete open game $\G : (X,S)\rightarrow (Y,R)$ using the 
diagram
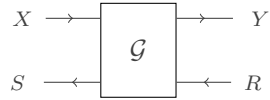
\begin{center}\tikzStart
	\node (g)[2morph]{$\G$};

\draw[->-] (g.40) -- +(0.75,0) node[label=east:{\fns$Y$}]{};

\path (g.-40) +(0.75,0) node (o2)[inner sep=0,label=east:{\fns$R$}]{};
\draw[->-] (o2) -- (g.-40);

\draw[->-] (g.220) -- +(-0.75,0) node[label=west:{\fns$S$}]{};

\path (g.140) +(-0.75,0) node (i1)[inner sep=0, label=west:{\fns$X$}]{};
\draw[->-] (i1) -- (g.140);

\tikzEnd\end{center}

This diagrammatic notation emphasises the point that information flows both covariantly
through $\G$ from observations to actions, and contravariantly through $\G$ from outcomes
to co-outcomes. These diagrams constitute a \textit{bona fide}
diagrammatic calculus for the category of concrete open games defined in the
remainder, as
detailed in \cite{hedges2017}.

\begin{notation}
	String diagrams in the category of open games will always be drawn with arrowheads
	on wires, whilst string diagrams in the ambient category will always be drawn
	without arrowheads. 
\end{notation}

\textit{Atomic} concrete open games are an important class of concrete open games, and are
the basic components out of which more complex games are constructed. Whilst concrete
open games can, in general, represent aggregates of agents responding
to each other (in a way that will be made precise in \ref{subsec:sComp} and
\ref{subsec:tComp}), atomic concrete open games describe games in which there is no
strategic interaction. Examples are simple computations in which no decisions are made
whatsoever, and single agents that are sensitive only to a given context. 

\begin{definition}[Atomic concrete open game]\label{atom}
	A concrete open game $a : (X,S)\rightarrow (Y,R)$ is \textit{atomic} if
	\begin{enumerate}
		\item $\Sigma_a\subseteq\conlens\big( (X,S),(Y,R)\big)$;
		\item For all $l\in\Sigma_a$, $\play_\G (l) = l$; and
		\item For all contexts $c : X\times (Y\rightarrow R)$, $\br_a(c) : \Sigma_a \to \pow (\Sigma_a)$ is
			constant.
	\end{enumerate}
	We sometimes refer to an atomic concrete open game simply as an \textit{atom}.
\end{definition}

Note that an atom $a : (X,S)\rightarrow (Y,R)$ is fully determined by a subset
$\Sigma_a \subseteq \conlens\big((X,S),(Y,R)\big)$ and a \textit{selection function}\footnote{Selection functions have been studied in a game-theoretic
  context in other form. See, e.g.~\cite{escardo2010selection} or \cite{hedges2017selection}.}
$\eps : X\times (Y\rightarrow R) \rightarrow \pow
(\Sigma_a)$.

Given $f : X\rightarrow Y$ and $g : R\rightarrow S$, as in $\conlens$, the pair
$(f,g)\in\set\times\set^{\op}$ can be represented as a concrete open game. We refer to
such games as \textit{computations}, as no strategic choice is being made.

\begin{example}[Computation]
	Let $f: X\rightarrow Y$ and $g:R\rightarrow S$. The atom
	$\langle f,g\rangle : (X,S)\rightarrow (Y,R)$ is given by
	\begin{enumerate}
		\item $\Sigma = \{\langle f,g\rangle\}$; and
		\item For all $c : X\times (Y\rightarrow R)$, $\eps (c) = \{ \langle
			f,g\rangle\}$.
	\end{enumerate}
\end{example}

Similar to $\conlens$, the following computations will turn out to be the underlying
structural maps for the symmetric monoidal category of concrete open games.

\begin{definition}[Structural computations]

Define identity, associator, swaps, and left/right unitor computations to be the
atomic concrete open games given by
\begin{align*}
\id_{(X,S)} &= \langle\id_X,\id_S\rangle
\\
\alpha_{X\otimes(Y\otimes Z),A\otimes (B\otimes C)} &=
\langle\alpha_{X,Y,Z},\alpha_{A,B,C}^{-1}\rangle
\\
s_{(X,A),(Y,B)} &= \langle s_{(X,Y)},s_{(A,B)}^{-1}\rangle
\\
\rho_{(X,Y)} &= \langle\rho_X,\rho_Y^{-1}\rangle
\\
\lambda_{(X,Y)} &= \langle\lambda_X,\lambda_Y^{-1}\rangle
\end{align*}
where the $\set$ functions on the right-hand side of the equalities are the obvious $\set$
isomorphisms.
\end{definition}

\textit{Counit games} are an interesting class of atoms that reverse the direction of information
flow in a concrete open game. 

\begin{definition}[Counit]
	Let $f : X\rightarrow S$. Define an atomic concrete open game $c_f :
	(X,S)\rightarrow (\iset,\iset)$ by
	\begin{enumerate}
		\item $\Sigma_{c_f} = \{ \langle !, f\rangle \}$; and
		\item For all $c : X\times (\iset\rightarrow\iset)$, $\eps(c) =
			\Sigma_{c_f}$.
	\end{enumerate}
\end{definition}

	We are being slightly relaxed with notation here as the update function for $c_f$
	has type $X\times\iset \rightarrow S$ while $f$ has type $X\rightarrow S$. We
	represent $c_f$ as follows.
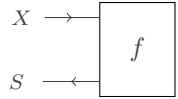
\begin{center}\tikzStart
	\node (f)[2morph]{$f$};

\path (f.140) +(-0.75,0) node (i1)[inner sep=0,label=west:{\fns$X$}]{};
\draw[->-] (i1) -- (f.140);

\draw[->-] (f.220) -- +(-0.75,0) node[label=west:{\fns$S$}]{};

\tikzEnd\end{center}

\subsection{Agents}\label{sec:conAgents}

So far we have only seen open games for which the set of strategies is a singleton,
describing games with no strategic decisions. Our first examples of a concrete open
game
with non-trivial strategy set are \textit{agents}. These can be used to represent the utility
maximising agents of traditional game theory or, more generally, to represent players
trying to influence the outcome of a game.

\begin{definition}[Agent]
	An \textit{agent} $\A : (X,\iset)\rightarrow (Y,R)$ is an atom whose set of strategies is $\Sigma =
	\conlens \big((X,\iset),(Y,R)\big)$.
\end{definition}

Recall that a concrete lens $l : \conlens \big( (X,\iset),(Y,R)\big)$ is a pair $( v: X\rightarrow Y,
u : X\times R\rightarrow \iset)$ and, hence, is uniquely determined by a function of type
$X\rightarrow Y$. Consequently, a \textit{strategy} for an agent specifies
how an agent map chooses an action of type $Y$ given an observation of type $X$. Given a context $c : X\times
(Y\rightarrow R)$, $\br_\A (c)$ picks out the set of strategies $\A$ considers acceptable
in the context $c$.
Agents are represented diagrammatically by

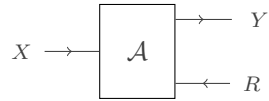
\begin{center}\tikzStart
	\node (a)[2morph]{$\A$};

\draw[->-] (a.40) -- +(0.75,0) node[label=east:{\fns$Y$}]{};
\path (a.-40) +(0.75,0) node (o2)[inner sep = 0, label=east:{\fns$R$}]{};
\draw[->-] (o2) -- (a.-40);

\path (a.west) +(-0.75,0) node(i)[inner sep = 0, label=west:{\fns$X$}]{};
\draw[->-] (i) -- (a.west);

\tikzEnd\end{center}

We can specialise the definition above to model the utility-maximising agents of
traditional game theory.

\begin{example}[Utility maximising agent]
	The \textit{utility maximising agent} $\A : (X,\iset)\rightarrow (Y,\R)$ is given
	by
	\[
		\eps(x,k) = \big\{ \sigma : X\rightarrow Y \bigm| \sigma (x)\in\arg\max(k)
		\big\}.
	\]
\end{example}

There are other decision criteria one could use. For instance, MinMax and regret minimisation would be candidates. We could also consider models from behavioural game theory such as prospect theory. The only (very weak) requirement is that the decision criterion can be described by a selection function \cite{hedges2017higher,hedges2017selection}. 
In this paper we focus on utility-maximising agents as a simplification and in order to match traditional game theory.

\subsection{Best-response with concrete lenses}\label{brLens}

Recall from \ref{conStates} and \ref{conEffects} that $\conlens\big( I, (X,S)\big) \cong
X$ and that $\conlens\big( I, (Y,R)\big) \cong Y\rightarrow R$. Using these facts we can
rephrase the type of best response for a concrete open game $\G : (X,S)\rightarrow (Y,R)$
as
\[
	\br_\G : \conlens\big( I, (X,S)\big) \times \conlens \big( (Y,R), I \big)
	\rightarrow \rel(\Sigma_\G).
\]

	This formulation allows for a concise and natural definition of \textit{sequential
	composition} for concrete open games where it would otherwise seem
	\textit{ad hoc}. To make matters clear, we write $x$ when talking about elements of
	$X$ and $x^\star$ when talking about concrete lenses with type $\conlens\big(
	I,(X,S)\big)$. Similarly, we write $k : Y\rightarrow R$ when talking about
	functions in $\set$ and we write $k_\star$ when talking about effects in $\conlens
	\big( (Y,R), I\big)$.

\subsection{Sequential composition of concrete open games}\label{subsec:sComp}

In this section we specify how to define the \textit{sequential
composite} $\HH\circ\G : (X,S)\rightarrow (Z,Q)$ of two concrete open
games $\G:(X,S)\rightarrow (Y,R)$ and $\HH: (Y,R)\rightarrow (Z,Q)$. 

We imagine that this composition \emph{really is} sequential in a
straightforward way. $\G$ is `played out' according to some strategy
$\sigma\in\Sigma_\G$ and then $\HH$ is `played out' according to some
$\tau\in\Sigma_\HH$. A choice of
$(\sigma,\tau)\in\Sigma_\G\times\Sigma_\HH$ therefore determines an open play
of $\G$ and $\HH$ played in sequence, and so we take $\Sigma_\G\times\Sigma_\HH$ to be the
set of strategy profiles of $\HH\circ\G$.

The play function of the sequential composite is defined straightforwardly using the
sequential composition of concrete lenses defined in \ref{seqclens}.

Defining best response for a sequential composite is a bit more
delicate and, for explanatory purposes, we make use of the informal
notion of a \textit{local context for a subgame}. Given a context
$c=(x:X, k:Z\rightarrow Q)$ and a strategy $(\sigma,\tau)$ for
$\HH\circ\G$, the best-response relation of $\HH\circ\G$
is specified by calling the best-response function of $\G$ with
a modified context corresponding to how $c$ `appears' to $\G$ when
$\HH$ plays according to $\tau$ and, similarly, calling the best-response function
of $\HH$ with a modified context corresponding to how $c$ `appears' to $\HH$ when $\G$ plays
according to $\sigma$. In practice we define these `local contexts' in the
obvious way that type checks, but this is because the work has already been done in
carefully choosing the correct definitions.

\begin{definition}[Sequential composition for concrete open games]
Let $\G = (\Sigma_\G, \play_\G, \br_\G) : (X,S)\rightarrow (Y,R)$ and $\HH = ( \Sigma_\HH,
\play_\HH, \br_\HH) : (Y,R)\rightarrow (Z,Q)$ be concrete open games. Define
\begin{enumerate}
\item $\Sigma_{\HH\circ \G} = \Sigma_\G \times \Sigma_\HH$,
\item $\play_{\HH\circ\G}(\sigma,\tau) = \play_\HH(\tau)\circ
\play_\G(\sigma)$ (where $\circ$ composition is in $\conlens$), and
\item $\br_{\HH\circ\G}(x^\star,k_\star)(\sigma,\tau) =
	\br_\G(x^\star,k_\star\circ\play_\HH(\tau))\sigma\times \br_\HH(\play_\G(\sigma)\circ
	x^\star,k_\star)(\tau)$.
\end{enumerate}
We represent $\HH\circ\G$ with the diagram
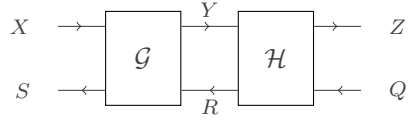
\begin{center}\tikzStart
	\input{tikz/scComp}
\tikzEnd \end{center}
\end{definition}

\subsection{Tensor composition for concrete open games}\label{subsec:tComp}

The tensor composition of open games represents simultaneous play. Given concrete open games $\G
:(X_1,S_1)\rightarrow (Y_1,R_1)$ and $\HH :(X_2,S_2)\rightarrow
(Y_2,R_2)$, the strategy set for $\G\otimes\HH$ is $\Sigma_\G\times\Sigma_\HH$; we make use of the tensor
composition in $\conlens$ in defining the play function; and the best-response function is given by
modifying the context $c$ to give local contexts for $\G$ and $\HH$. 

\begin{definition}[Local contexts for tensor composition]
	Define the \textit{left local tensor context operator}
	\[
		\leftc : \Big( X' \times (X'\rightarrow Y') \times (Y\times Y'\rightarrow
		R\times R')\Big) \rightarrow (Y\rightarrow R)
	\]
	by
	\[
		\leftc (x',p',k)(y) = \pi_1 \circ k(y,p'(x')).
	\]
	As a diagram, $\leftc(x',p',l)$ is the function
	\begin{center}\tikzStart
		\input{tikz/conLocal1}.
	\tikzEnd\end{center}
	Similarly, define the \textit{right local tensor context operator}
	\[
		\rightc : \Big( X \times (X\rightarrow Y) \times (Y\times Y'\rightarrow
		R\times R')\Big) \rightarrow (Y'\rightarrow R')
	\]
	by
	\[
		\rightc (x,p,k)(y') = \pi_2 \circ k(p(x), y).
	\]
	As a diagram,
	\begin{center}\tikzStart
		\input{tikz/conLocal2}.
	\tikzEnd\end{center}
\end{definition}

Suppose we have concrete open games $\G : (X,S)\rightarrow (Y,R)$ and $\HH :
(X',S')\rightarrow (Y',R')$ and we wish to combine them to create some game $\G\otimes \HH
: (X\times X',S\times S')\rightarrow (Y\times Y', R\times R')$. Consider the left context
operator $\leftc$ acting on some triple $(x',p',k)$. If $k$ is an outcome function for the
game $\G\otimes\HH$ and $\HH$ observes $x'$ and plays according to the function $p'$, then
$\leftc(x',p',k)$ is the `apparent' outcome function for $\G$. Similarly, $\rightc(x,p,k)$
is the `apparent' outcome function for $\HH$ when $\G$ observes $x$ and plays according to
$p$. With this in mind, we define tensor composition for concrete open games as follows.

\begin{definition}[Tensor composition of concrete open games]
Let $\G : (X,S)\rightarrow (Y,R)$ and $\HH : (X',S')
\rightarrow (Y',R')$ be concrete open games. Define
\[
\G\otimes\HH : (X\times X', S\times S') \rightarrow (Y\times
Y', R\times R')
\]
by
\begin{enumerate}
\item $\Sigma_{\G\otimes\HH} = \Sigma_\G \times \Sigma_\HH$,
\item $\play_{\G\otimes\HH}(\sigma,\tau) = \play_\G (\sigma) \otimes
	\play_\HH (\tau)$ (in $\conlens$),
\item $\br_{\G\otimes\HH}: \Big( (X\times X') \times (Y\times Y'\rightarrow R\times R')\Big)
	\rightarrow \rel(\Sigma_{\G\otimes\HH})$ is given by
	\begin{align*}
		\br_{\G\otimes\HH}\big((x,x')^\star ,k_\star\big)(\sigma,\tau) &= \br_\G
		\big(x^\star,
		\leftc(x',(\play_\HH(\tau))_v,k)_\star\big)(\sigma) \\
									      &\quad \times \br_\HH
		\big(x'^\star,\rightc(x,(\play_\G(\sigma))_v, k)_\star\big)(\tau)
	\end{align*}
\end{enumerate}
$\G\otimes\HH$ is represented by the diagram
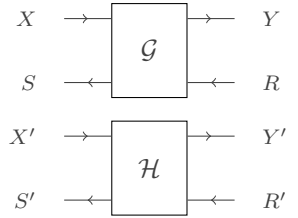
\begin{center}\tikzStart
	\input{tikz/tcComp}
\tikzEnd \end{center}
\end{definition}

\subsection{Equivalence of open games}\label{iso}

One subtlety remains before we can define the category of concrete open games. 
We aim to define a category with pairs of sets as objects
and morphisms given by concrete open games. If carried out na\"ively, this runs into the
problem that strategy sets which should be identical are merely isomorphic. For instance,
the strategy set of $\K\circ(\HH\circ\G)$ is
$(\Sigma_\G\times\Sigma_\HH)\times\Sigma_\K$
whilst the strategy set of $(\K\circ\HH)\circ\G$ is $\Sigma_\G \times
(\Sigma_H\times\Sigma_\K)$. In order for concrete open games to form a category, we must
first take an appropriate quotient. 

There are several different reasonable choices of quotient. Since this is an orthogonal consideration to this paper's topic, we choose the most straightforward, which is to identify open games that have a compatible isomorphism between their sets of strategies. Other choices that can be made are bisimulations \cite{bolt_probability_nondeterminism} and surjections \cite{fong-spivak-tuyeras-backprop-as-functor}. Alternatively, instead of taking a quotient, open games can be considered as the 1-cells of a bicategory \cite{morphisms}.



\begin{definition}
	Let $\G,\HH : (X,S)\rightarrow (Y,R)$ be concrete open games. An \textit{isomorphism}
	$\alpha : \G\rightarrow \HH$ is given by a bijection $\alpha : \Sigma_\G
	\to \Sigma_\HH$ such that
	\begin{enumerate}
		\item $\play_\G (\sigma) = \play_\HH (\alpha (\sigma))$ for all $\sigma \in \Sigma_\G$
		\item For all $\sigma, \sigma' \in \Sigma_\G$ and $c \in X \times (Y \to R)$, $(\sigma, \sigma') \in \br_\G (c)$ iff $(\alpha (\sigma), \alpha (\sigma')) \in \br_\HH (c)$
	\end{enumerate}
\end{definition}


\begin{definition}
	Let $\G,\HH : (X,S)\rightarrow (Y,R)$ be concrete open games. $\G$ and $\HH$ are
	\textit{equivalent}, written $\G\sim\HH$, if there exists an isomorphism $\alpha :
	\G\rightarrow \HH$. We write $[\G]$ for the equivalence class of $\G$ under this
	relation. We also say that the isomorphism $\alpha$ \textit{witnesses} the
	equivalence between $\G$ and $\HH$ and write $\G\overset{\alpha}{\sim}\HH$.
\end{definition}


The following results demonstrate that sequential and tensor composition of concrete open
respects equivalence of concrete open games.

\begin{lemma}
	Let $\G,\G' : (X,S)\rightarrow (Y,R)$ and $\HH,\HH' : (Y,R)\rightarrow (Z,Q)$ be
	concrete open games. If $\G\sim\G'$ and $\HH\sim\HH'$, then
	$\HH\circ\G\sim\HH'\circ\G'$.
\end{lemma}

\begin{proof}
	Suppose $\G\overset{\alpha}{\sim}\HH$ and $\G'\overset{\beta}{\sim}\HH'$. Then
	$\alpha \times \beta : \Sigma_\G \times \Sigma_\G \to
	\Sigma_{\G'}\times\Sigma_{\HH'}$ given by
	\[ (\alpha \times \beta) (\sigma, \tau) = (\alpha (\sigma), \beta (\tau)) \]
	is such that $\HH\circ\G \overset{\alpha\times\beta}{\sim} \HH'\circ\G'$.
\end{proof}

\begin{lemma}
	Let $\G, \HH: (X,S)\rightarrow (Y,R)$ and $\G', \HH' : (X',S')\rightarrow (Y',R')$
	be concrete open games. If $\G\sim\HH$ and $\G'\sim\HH'$, then
	$\G\otimes\G'\sim\HH\otimes\HH'$.
\end{lemma}

\begin{proof}
 If $\G\overset{\alpha}\sim \HH$ and $\G'\overset{\beta}{\sim}\HH'$, then $\G\otimes\G'
 \overset{\alpha\times\beta}{\sim} \HH\otimes\HH'$ as in the previous lemma.
\end{proof}


\subsection{The category of concrete open games}\label{subsec:conCat}

We are now finally in a position to show that concrete open games form a symmetric
monoidal category.

\begin{notation}
	In string diagrams we refer to a play function applied to a strategy simply by the
	strategy. For example, $\sigma$ may refer to $\play_\G(\sigma)$. In practice this
	does not lead to ambiguity because proofs and definitions proceed by assigning fixed
	strategies to particular open games. This notational convention allows for less
	cluttered string diagrams.
\end{notation}

\begin{lemma}
Sequential composition of concrete open games is
associative up to equivalence.
\end{lemma}

The identity morphism $(X,S)\rightarrow (X,S)$ is given by the computation $\langle\id_X,
\id_S\rangle$.

\begin{lemma}
	Let $\G : (X,S)\rightarrow (Y,R)$. Then $[\G] = [\G\circ
	\langle\id_X,\id_S\rangle]=[\langle\id_Y,\id_R\rangle\circ\G]$.
\end{lemma}

\begin{corollary}
	There is a category $\congame$ with pairs of sets as object
	and equivalence
	classes of concrete open games as morphisms.\qed
\end{corollary}

We now move on to proving that $\congame$ is symmetric monoidal.

\begin{lemma}
$\otimes : \congame\times\congame\rightarrow\congame$ is a functor.
\end{lemma}

\begin{lemma}
The associator in $\congame$ is natural.
\end{lemma}

We include the proof of this lemma specifically because it will be important later.

\begin{proof}
Let $\G_i : (X_i,S_i) \rightarrow (Y_i,R_i)$ be open games where $i\in\{1,2,3\}$. We need
to show that
$\alpha\circ(\G_1\otimes(\G_2\otimes\G_3))\sim((\G_1\otimes\G_2)\otimes\G_3)\circ\alpha$.
Define $\beta :
(\Sigma_{\G_1}\times(\Sigma_{\G_2}\times\Sigma_{\G_3}))\times\iset\to \iset\times
((\Sigma_{\G_1}\times\Sigma_{\G_2})\times\Sigma_{\G_3})$ by
$\beta \big((\sigma,(\tau,\mu)),\star\big) = \big(\star,
((\sigma,\tau),\mu)\big) $.

As $\conlens$ is symmetric monoidal, we have that
\[\alpha\circ\play_{\G_1\otimes(\G_2\otimes\G_3)}(\sigma,(\tau,\mu)) =
\play_{(\G_1\otimes\G_2)\otimes\G_2}((\sigma,\tau),\mu)\circ\alpha.\]

Let $x_i\in X_i$ and $k : (Y_1\times Y_2\times Y_3)\rightarrow (R_1\times R_2\times R_3)$.
We will show that the local contexts for $\G_1, \G_2,$ and $\G_3$ are the same in both
$\G_1\otimes(\G_2\otimes\G_2)$ and $(\G_1\otimes\G_2)\otimes\G_3$. First we consider
$\G_1$. Let $k' := \leftc (x_3, \play_{\G_3}\mu,k)$. Then,
	\[
		k_{\G_1}:=
		\leftc\big( (x_2,x_3), \play_{\G_2\otimes\G_3}(\tau,\mu),
		k\big) 	\]
	
		
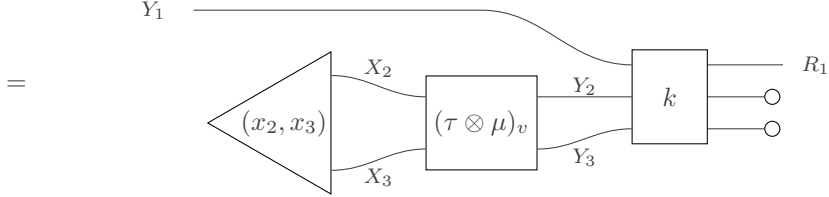
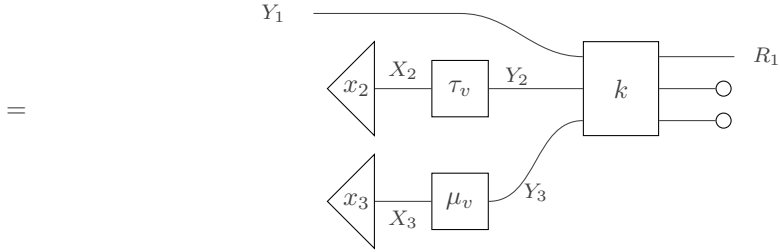
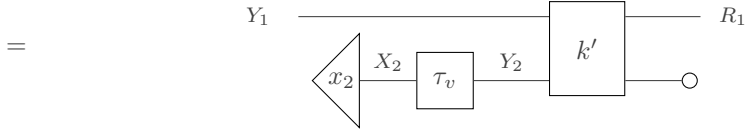
\vspace{0.5cm}

		\rb{40pt}{$=$\quad\quad\quad\quad}\tikzStart\input{tikz/cGameA4}\tikzEnd

	\vspace{1cm}

	\rb{50pt}{$=$\hspace{3cm}}\tikzStart\input{tikz/cGameA5}\tikzEnd

	\vspace{15pt}

	\rb{1cm}{$=$\hspace{2.8cm}}\tikzStart\input{tikz/cGameA6}\tikzEnd

	\vspace{0.5cm}

	\[
		= \leftc\Big( x_2, \play_{\G_2}(\tau),
		\leftc\big(x_3,\play_{\G_3}\mu,k\big)\Big)
	\]
Similar arguments hold for $\G_2$ and $\G_3$, showing that
\[
	k_{\G_2} := \rightc\Big( x_1, \play_{\G_1}(\sigma), \leftc\big(
			x_3,\play_{\G_3}(\mu),
	k\big)\Big)
	=
	\leftc\Big( x_3, \play_{\G_3} (\mu), \rightc\big( x_1,\play_{\G_1}(\sigma),
	k\big)\Big)
\]
and
\[
	k_{\G_3} := \rightc\Big( (x_1,x_2), \play_{\G_1\otimes\G_2}(\sigma,\tau), k \Big)
	=
	\rightc\Big( x_2, \play_{\G_2}(\tau), \rightc\big( x_1, \play_{\G_1}(\sigma),
	k\big)\Big).
\]
Then
\begin{align*}
	&\big((\star, (\sigma, (\tau, \mu))), (\star, (\sigma', (\tau', \mu')))\big) \in \br_{\alpha\circ(\G_1\otimes(\G_2\otimes\G_3))}\big((x_1, (x_2,x_3)), k\big) \\
	\iff\ &(\sigma, \sigma') \in \br_{\G_1}(x_1, k_{\G_1}) \text{ and } (\tau, \tau') \in \br_{\G_2}(x_2, k_{\G_2}) \text{ and } (\mu, \mu') \in \br_{\G_3}(x_3, k_{\G_3}) \\
	\iff\ &\big((((\sigma, \tau), \mu), \star), (((\sigma', \tau'), \mu'), \star)\big) \in \br_{((\G_1\otimes\G_2)\otimes\G_3)\circ\alpha}\big(((x_1,x_2),x_3), k \big)
\end{align*}
%
\end{proof}

The above lemma relies on the fact that the monoidal tensor in $\set$
is cartesian. In particular we needed that bipartite states $s :
I\rightarrow S_1\otimes S_2$ in $\set$ (i.e.~elements of $S_1\times S_2$)
correspond to pairs of states $(s_1 : I\rightarrow S_1, s_2 :
I\rightarrow S_2)$. In an arbitrary monoidal category, it need not be
the case that for all states $s: I \rightarrow S_1\otimes S_2$ there
exist states $s_1 : I\rightarrow S_1$ and $s_2 : I \rightarrow S_2$
such that
\begin{center}
\rb{18pt}{\tikzStart\node (s) [2state]{$s$};

\path (s.60) +(0.5,0) node (o1)[label=east:{\fns$S_1$}]{};
\draw (s.60) -- (o1);

\path (s.-60) +(0.5,0) node (o2)[label=east:{\fns$S_2$}]{};
\draw (s.-60) -- (o2);
\tikzEnd}
\rb{32pt}{$\quad=\quad$}
\tikzStart\node (s1) [2state] {$s_1$};
\path (s1.east) +(0.5,0) node (o1)[label=east:{\fns$S_1$}]{};
\draw (s1.east) -- (o1);

\path (s1.south) +(0,-1) node (s2)[2state]{$s_2$};
\path (s2.east) +(0.5,0) node (o2)[label=east:{\fns$S_2$}]{};
\draw (s2.east) -- (o2);
\tikzEnd
\end{center}
This poses a significant barrier to generalising concrete open games
to monoidal categories where the monoidal tensor is not cartesian, and
Section \ref{chap:ogames} addresses this problem.

\begin{lemma}
The structural computations $\lambda, \rho,$ and $s$ are natural in
$\congame$.
\end{lemma}

\begin{theorem}
	$\congame$ is symmetric monoidal.
\end{theorem}

\subsection{Encoding functions as games}\label{functions}

Recall that, given functions $f:X\rightarrow Y$ and $g : R\rightarrow S$, there is a
computation of concrete open games $\langle f,g\rangle : (X,S)\rightarrow (Y,R)$. In fact,
this operation is functorial.

\begin{lemma}[\cite{jules}]
	Define $F : \set\times\set^\op \rightarrow \congame$ by $F(X,S) = (X,S)$ and $F(
	f:X\rightarrow Y, g:R\rightarrow S) = \langle f,g\rangle$. Then $F$ is a faithful
	monoidal functor.\qed
\end{lemma}

We also incorporate computations directly into the diagrammatic calculus for concrete open
games, representing the computation $\langle f,g\rangle : (X,S)\rightarrow (Y,R)$ by
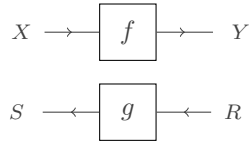
\begin{center}
	\tikzStart\input{tikz/computationDiagram}\tikzEnd 
\end{center}
Two particularly useful examples of this notation are the covariant and contravariant
copying computations $\langle \Delta_X, \id_1 \rangle : (X,1) \rightarrow (X\times X, 1)$
and $\langle \id_1, \Delta_R\rangle : (1,R\times R) \rightarrow (1,R)$ which are represented
by
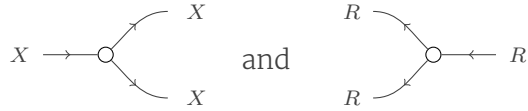
\begin{center}
	\tikzStart\node (c)[copy]{};

\path (c.west) +(-0.75,0) node (i)[inner sep=0,label=west:{\fns$X$}]{};
\draw[->-] (i) -- (c.west);

\draw[->-] (c.45) to [out=45,in=180] +(0.75,0.5) node[label=east:{\fns$X$}]{};
\draw[->-] (c.-45) to [out=-45,in=180] +(0.75,-0.5) node[label=east:{\fns$X$}]{};
\tikzEnd
	\rb{17pt}{\quad and \quad}
	\tikzStart\node (c)[copy]{};

\path (c.135) +(-0.75,0.5) node (i1)[inner sep=0,label=west:{\fns$R$}]{};
\path (c.225) +(-0.75,-0.5) node (i2)[inner sep=0,label=west:{\fns$R$}]{};
\draw[->-] (c.135) to [out=135,in=0] (i1);
\draw[->-] (c.225) to [out=225,in=0] (i2);

\path (c.east) +(0.75,0) node (o)[inner sep=0,label=east:{\fns$R$}]{};
\draw[->-] (o) -- (c.east);
\tikzEnd
\end{center}
respectively.
\subsection{Game theory with concrete open games}\label{sec:conTheory}

In this section we give some examples of games modelled using concrete open games. We will
be light on details, aiming to simply demonstrate some of the expressive power of concrete
open games. We direct the reader to \cite{jules} for more details.

\subsubsection{Bimatrix games}\label{conBimatrix}

Bimatrix games are simply two-player normal-form games, the most well-known example of
which is likely the prisoner's dilemma. We assume
the set of actions available to each player is finite for simplicity.

\begin{definition}
	A \textit{bimatrix game} consists of
	\begin{enumerate}
		\item Finite set of actions $A$ and $B$; and
		\item An outcome function $k : A\times B \rightarrow \R^2$.
	\end{enumerate}
\end{definition}
A bimatrix game $\G = (A,B,k)$ is represented by the concrete open game
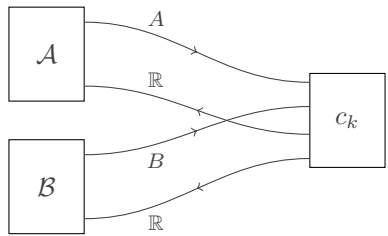
\begin{center}
	\tikzStart\input{tikz/bimatrix}\tikzEnd
\end{center}
where $\A$ and $\bAgent$ are utility maximising agents and $c_k$ is the counit game
associated with $k$. In diagrammatic form, the structure of the game is made clear. Players
$\A$ and $\bAgent$ make independent choices from $A$ and $B$ respectively which are then used
to generate two real numbers as outcomes. Bimatrix games may not have a Nash equilibrium
in pure strategies, but in cases that do have Nash equilibria, they appear as fixed points
of the best-response function $\br : A \times B \to \pow (A \times B)$ of 
the above concrete open game, i.e.~strategy profiles $(a, b)$ satisfying $(a, b) \in \br (a, b)$.

\subsubsection{Two-player sequential game}\label{subsec:twoPlay}

A two-player sequential game is defined by the same data as a bimatrix game (sets $A$ and $B$ and a function $k : A \times B \to \R^2$), but we allow the second player to observe the first player's move before making a choice, so strategies for the second player are functions $A \to B$.
This is represented by the concrete open game
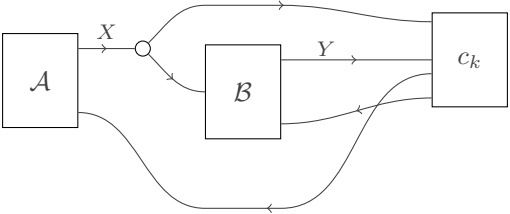
\begin{center}
	\tikzStart\input{tikz/openSequential}\tikzEnd
\end{center}
where $\A$ and $\bAgent$ are utility maximising agents and $c_k$ is the counit game associated with $k$.

Crucially, the fixed points of the concrete open game are \emph{not} subgame perfect Nash
equilibria, but rather plain old Nash equilibria.  
It is also possible to define a concrete open game that captures subgame perfect equilibria, but this requires an additional operator defined in \cite{ghani}.

\subsubsection{Normal-form games}\label{normalform}

Let $\Gamma = \big(N , (S_i)_{i=1}^N, (u_i)_{i=1}^N \big)$ be a normal-form game
for $N$ players. $S_i$ denotes the set of strategies available to player $i$.
The function $u_i$ maps a strategy tuple for all
players, $ \prod_{i=1}^N S_i$, to player i's payoff, in $\R$.   
Define $k : \prod_{i=1}^N S_i \rightarrow \R^N$ by $s=(s_1,\cdots, s_N)\mapsto
(u_1(s),\cdots, u_N(s))$. We can model this normal-form game using the concrete open game
\[
	c_k \circ \Big(\bigotimes_{i=1}^N \A_i \Big)
\]
where $\A_i : I \rightarrow (S_i, \R)$ is the utility maximising agent. The fixed points
of this game's best-response relation are then the pure-strategy Nash equilibria of the normal-form game.

\section{General open games}
\label{chap:ogames}

The notion of open game we introduced in the section before can emulate some
standard games such as the prisoner's dilemma. On the other hand, classical game
theory has a much wider reach. It can model situations with which 
a concrete open game cannot deal. This involves stochastic environments,
probabilistic choices by players, and incomplete information. 

In this section, we will significantly generalise the notion of an open game,
to make room for these three extensions (and beyond).  The first order of
business, to make progress in this direction, is to generalise concrete lenses.

\subsection{Generalising concrete lenses}\label{sec:genLens}

In the proof of Lemma~\ref{cLensAssoc} we made use of the fact that every $\set$ function is a
comonoid homomorphism for the copy/delete comonoid. Recall that a morphism is a comonoid
homomorphism if it can be `moved through' the comonoid structure.
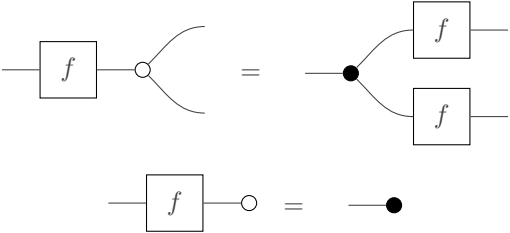
\begin{center}
	\rb{12pt}{\tikzStart\node (f) [1morph]{$f$};
\draw (f.west) -- +(-0.5,0);
\draw (f.east) -- +(0.5,0) node (c)[copy,anchor=west]{};
\draw (c.45) to [out=45,in=180] +(0.75,0.5);
\draw (c.-45) to [out=-45,in=180] +(0.75,-0.5);
\tikzEnd}
	\rb{25pt}{\quad $=$ \quad}
	\tikzStart\node (c)[varCopy]{};

\draw (c.45) to [out=45,in=180] +(0.75,0.5) node (f1)[1morph,anchor=west]{$f$};
\draw (c.-45) to [out=-45,in=180] +(0.75,-0.5) node (f2)[1morph,anchor=west]{$f$};

\draw (f1.east) -- +(0.5,0);
\draw (f2.east) -- +(0.5,0);
\draw (c.west) -- +(-0.5,0);
\tikzEnd

	\vspace{10pt}

	\tikzStart\node (f) [1morph]{$f$};
\draw (f.east) -- +(0.5,0) node (c)[copy,anchor=west]{};
\draw (f.west) -- +(-0.5,0);
\tikzEnd
	\rb{7pt}{\quad $=$ \quad}
	\rb{7pt}{\tikzStart\node (c)[varCopy]{};
\draw (c.west) -- +(-0.5,0);
\tikzEnd}
\end{center}

If $\set$ is replaced with some
arbitrary symmetric monoidal category $\C$ and the copy/delete comonoid is replaced with
some arbitrary comonoid in $\C$, sequential composition of lenses, as defined in Definition
\ref{seqclens}, may not be
associative. This presents a substantive problem --- there exist categories relevant to
game theory in which sequential composition of concrete lenses is not associative. Of
particular interest is the Kleisli category of the finitary distribution monad, $\kl
(D)$, which we will need in order to model Bayesian games (discussed in Section 
\ref{chap:bGames}). $\kl(D)$ inherits a copy/delete
comonoid from $\set$, but its comonoid homomorphisms are the deterministic 
maps (i.e.~precisely the non-probabilistic maps).

In the next section we introduce \textit{coends}, a piece of categorical machinery that
allows for an elegant generalisation of concrete lenses to arbitrary symmetric monoidal
categories. We call these generalised lenses \textit{coend lenses} or, simply, \textit{lenses}. We will first introduce the technical notion before
magicking it away with a diagrammatic calculus that represents what is `really' going on.

\subsection{Co-wedges and Coends}\label{sec:coends}

Co-wedges are a variant of co-cones of
natural transformations applying to functors that act both covariantly and contravariantly
on an argument. In Section~\ref{sec:lens} we noted
that lenses have both covariant and contravariant components. We will see that this behaviour can be
described by coends, which are initial co-wedges. For extra motivation, discussion, and
examples, we refer the reader to~\cite{loregian}. 

\begin{definition}[Co-wedge]
Let $F:\C^{\op}\times\C \rightarrow \D$ be a functor. A
\textit{co-wedge} $c:F\rightarrow\alpha$ is an object $\alpha:\D$
together with maps $\big\{ c_a : F(a,a)\rightarrow \alpha \bigm| a : \C \big\}$ such that, for
any morphism $f:a'\rightarrow a$, the diagram

\begin{center}\tikzStart
\node (tl) {$\alpha$};
\path (tl.east) +(3,0) node (tr){$F(a,a)$} [anchor=west];
\path (tl.south) +(0,-1) node (bl)[anchor=north] {$F(a',a')$};
\path (tr |- bl) node (br) {$F(a',a)$};

\draw[<-] (tl.east) -- node[above]{\fns$c_a$} (tr.west);
\draw[<-] (tl.south) -- node[left]{\fns$c_{a'}$} (bl.north);
\draw[<-] (bl.east) -- node[below]{\fns$F(a',f)$} (br.west);
\draw[<-] (tr.south) -- node[right]{\fns$F(f,a)$} (br.north);
\tikzEnd\end{center}
commutes.
\end{definition}

\begin{definition}[Coend]
A \textit{coend} is a couniversal co-wedge. Diagrammatically, the coend of a functor $F
:\C^{\op}\times\C\rightarrow D$ is a co-wedge $\big\{ c_a:F(a,a) \rightarrow
	\mathrm{coend}(F) \bigm| a : \C
\big\}$ such that for any other co-wedge $\big\{ d_a : F(a,a)
\rightarrow \alpha \bigm| a : \C \big\}$ and morphism $f:a'\rightarrow a$ the
diagram

\begin{center}\tikzStart
\node (tl)  {$\mathrm{coend}(F)$};
\path (tl.east) +(3,0) node (tr)[anchor=west]{$F(a,a)$};
\path (tl.south) +(0,-1) node (bl)[anchor=north]{$F(a',a')$};
\path (tr |- bl) node (br){$F(a',a)$};
\path (tl.north west) +(-1,1) node (tll) {$\alpha$};

\draw[->] (br.north) -- node[right]{\fns$F(f,a)$} (tr.south);
\draw[->] (br.west) -- node[below]{\fns$F(a',f)$} (bl.east);
\draw[->] (tr.west) -- node[above]{\fns$c_a$} (tl.east);
\draw[->] (bl.north) -- node[left]{\fns$c_{a'}$} (tl.south);
\draw[->] (tr.north west) to [out=135,in=0,out looseness=0.5] 
	node[above]{\fns$d_a$} (tll.east);
\draw[->] (bl.north west) to [out=135,in=-90,out looseness=0.5]
	node[left]{\fns$d_{a'}$} (tll.south);
\draw[->,dashed] (tl.north west) -- node[right]{\fns$h$} (tll.south east);
\tikzEnd\end{center}
commutes for a unique morphism $h:\mathrm{coend}(F) \rightarrow \alpha$.
\end{definition}
We adopt the integral notation for coends, writing 
\[
\int^{a:\C}F(a,a)
\]
for $\mathrm{coend}(F)$. We will make use of the fact that coends can be characterised by
the following coequaliser.

\begin{lemma}\label{coeqCoend}

	Let $F : C^{\op}\times \C \rightarrow \D$. If $\D$ is cocomplete and $\C$ is
	small, the coend $\int^{a:\C}F(a,a)$ is given by the coequaliser of the 
	pair of arrows
\begin{center}\tikzStart
\node (l) {$\displaystyle\coprod_{\substack{a,a':\C \\
f:a'\rightarrow a}} F(a,a')$};
\path (l.north east) +(2,0) node (r)[anchor=north
west]{$\displaystyle\coprod_{a:\C}F(a,a)$,};

\draw[->] (l.17) -- node[above]{\fns$F_1$} (r.west |- l.17);
\draw[->] (l.10) -- node[below]{\fns$F_2$} (r.west |- l.10);
\tikzEnd\end{center}
where the $f:a'\rightarrow a$ components of $F_1$ and $F_2$ are
$F(f,a')$ and $F(a,f)$ respectively.
\qed
\end{lemma}
When $\C$ is \emph{not} small (as it usually is not), we need to show directly that
coends exist.

\subsection{Coend lenses}\label{sec:coendLens}

Much of the material in this section is worked out in much greater detail in~\cite{mitchell},
which serves as a good standard reference for coend lenses.
We first give an abstract definition of coend lenses, then provide some justification.

\begin{definition}[Coend lens]
Let $X,S,Y,$ and $R$ be objects in a symmetric monoidal category $\C$.
A \textit{coend lens}  $l:(X,S)\rightarrow (Y,R)$ is an element of the set
\[
\int^{A:\C} \C(X,A\otimes Y) \times \C(A\otimes R,S)  .
\]
\end{definition}

We think of the coend in the above definition as acting as a kind of existential
quantifier over the type variable $A$, followed by a quotient (to be described) over the resulting structure. That is, a coend lens $l : (X,S)\rightarrow (Y,R)$
consists of an equivalence relation over triples comprised of a choice of type $A$, a
morphism $v : X\rightarrow A\otimes Y$, and another morphism $u : A\otimes R\rightarrow
S$.  

By Lemma \ref{coeqCoend} we can characterise coend lenses $(X,S)\rightarrow (Y,R)$ as the
elements of a particular coequaliser. Moreover, coequalisers in $\set$ are given by
quotients. Unpacking the coequaliser explicitly, coend lenses $(X,S)\rightarrow (Y,R)$
are given by the set of triples of the form described above, quotiented by the equivalence
relation generated by (i.e.~the smallest equivalence relation containing)
\[
\big( (f\otimes\id_Y)\circ v,u\big) \sim \big(v,
u\circ(f\otimes\id_R)\big)
\]
for all $A,B:\C$ and $f :A\rightarrow B$.
In diagrammatic form, the pair
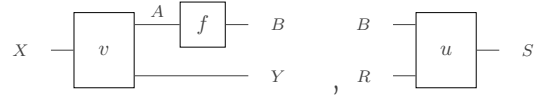
\begin{center}
\scalebox{0.8}{\tikzStart\input{tikz/endLens1}\tikzEnd} , 
\scalebox{0.8}{\tikzStart\node (u) [2morph] {$u$};

\path (u.east) +(0.5,0) node (o1)[label=east:{\fns$S$}]{};
\path (u.140) +(-0.5,0) node (i1)[label=west:{\fns$B$}]{};
\path (u.220) +(-0.5,0) node (i2)[label=west:{\fns$R$}]{};

\draw (u.east) -- (o1);
\draw (u.140) -- (i1);
\draw (u.220) -- (i2);

\path (o1) +(-1,0) node (buffer){};
\tikzEnd}
\end{center}
is related to the pair
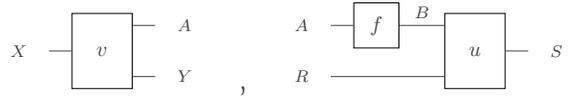
\begin{center}
\scalebox{0.8}{\tikzStart\node (v) [2morph]{$v$};

\path (v.west) +(-0.5,0) node (i1)[label=west:{\fns$X$}]{};
\path (v.40) +(0.5,0) node (o1)[label=east:{\fns$A$}]{};
\path (v.-40) +(0.5,0) node (o2)[label=east:{\fns$Y$}]{};

\draw (v.west) -- (i1);
\draw (v.40) -- (o1);
\draw (v.-40) -- (o2);

\path (o1) +(1,0) node (buffer){};
\tikzEnd} ,
\scalebox{0.8}{\tikzStart\input{tikz/endLens4}\tikzEnd}
\end{center}
We refer to the types $A$ and $B$ as \textit{bound types} ($B$ is bound in the first
diagram, $A$ in the second). In Section \ref{chap:bGames}, we will see that this bound type
keeps track of \textit{correlations} between random variables in the Kleisli category of
the distribution monad.

In vague terms, two pairs of morphisms are related if one
can get from one to the other by `sliding' a morphism off the bound type of one morphism
 on to the bound type of the other. Given a pair of morphisms $(v :
X\rightarrow A\otimes Y, u : A\otimes R\rightarrow S)$, we write $[v,u]$ for their
equivalence class. When we need to talk explicitly about the bound type of $[v,u]$ we write
$[A,v,u]$ to specify that the pair $(v,u)$ has bound type $A$. We also adopt the
convention that $ l = [A_l, l_v, l_u]$ where, as with concrete lenses, we say that $l_v$
is the \textit{view morphism} and $l_u$ is the \textit{update morphism}.
We follow \cite{mitchell}, taking the hint from the diagrammatic representation of the
equivalence relation by representing a coend lens $[v,u]
:(X,S) \rightarrow (Y,R)$ as
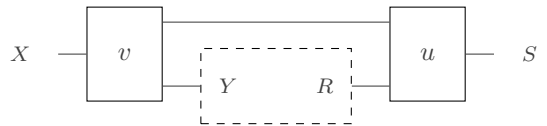
\begin{center}
\tikzStart\input{tikz/coLens}\tikzEnd
\end{center}
We usually omit the bound type in diagrams for clarity.
The equivalence relation is then simply
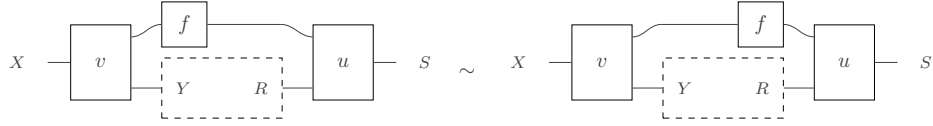
\begin{center}
\scalebox{0.8}{
\tikzStart\input{tikz/coLens2}\tikzEnd
\rb{20pt}{$\quad \sim \quad$}
\tikzStart\input{tikz/coLens3}\tikzEnd}
\end{center}
The equivalence relation permits the cancelling of isomorphisms:
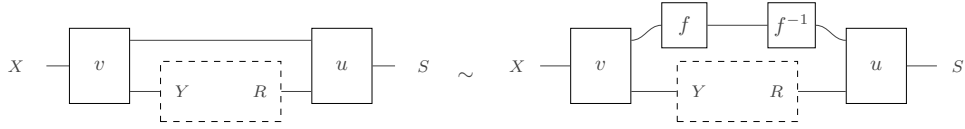
\begin{center}
	\scalebox{0.8}{
	\tikzStart\input{tikz/coLens}\tikzEnd
	\rb{20pt}{$\quad \sim \quad$}
\tikzStart\input{tikz/lensCancel}\tikzEnd}
\end{center}
Many proofs in this section proceed by allowing symmetric monoidal structure to interact
with coend structure as, for example, in the following diagram.
\begin{center}
	\scalebox{0.8}{
	\tikzStart\input{tikz/lensSwap1}\tikzEnd
	\rb{20pt}{$\quad \sim \quad$}
\tikzStart\input{tikz/lensSwap2}\tikzEnd}
\end{center}

The formal foundations of this class of diagrams are investigated in \cite{roman_open_diagrams}.

\begin{example}[Identity lens]
The \textit{identity lens} $\id_{(X,S)} : (X,S)\rightarrow (X,S)$ is
given by $[I, \id_X : X\rightarrow X, \id_S : S\rightarrow S]$. Diagrammatically,
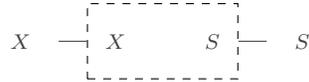
\begin{center}
\tikzStart\node (g)[dbox]{};

\path (g.west) +(-0.5,0) node (i)[label=west:{\fns$X$}]{};
\path (g.west) node [label=east:{\fns$X$}]{};
\draw (i) -- (g.west);

\path (g.east) +(0.5,0) node (o)[label=east:{\fns$S$}]{};
\path (g.east) node [label=west:{\fns$S$}]{};
\draw (g.east) -- (o);
\tikzEnd
\end{center}
\end{example}

\begin{example}
A pair of morphisms $(f:X\rightarrow Y, g: R\rightarrow S)$ is
encoded  by the coend lens $[I, f,g] :(X,S)\rightarrow (Y,R)$:
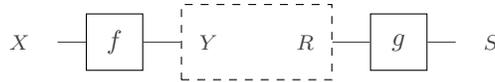
\begin{center}
\tikzStart\input{tikz/morphEmbed}\tikzEnd
\end{center}
\end{example}

\begin{definition}[Sequential composition of coend lenses]
Let $[v,u]:(X,S)\rightarrow (Y,R)$ and $[v',u'] :(Y,R)\rightarrow
(Z,Q)$ be coend lenses. Define $[v',u']\circ[v,u] : (X,S)\rightarrow
(Z,Q)$ to be
\begin{center}
\tikzStart\input{tikz/compCoLens1}\tikzEnd
\end{center}
Explicitly,
\[
	[A',v',u']\circ[A,v,u] = [A\otimes A',(v'\otimes\id_A)\circ v, u\circ(\id_A\otimes u')]. 
\]
\end{definition}

\begin{theorem}[Coend lenses form a category]
Suppose $\C$ is a monoidal category such that, for all objects $X,S,Y,R \in \C$,
\[
	\int^{A:\C} \C (X,A\otimes Y) \times \C(A\otimes R,S)	
\]
exists. Then there is a category $\lens_\C$ whose objects are pairs of 
objects in  $\C$ and where
\[
	\lens_\C \big((X,S),(Y,R)\big) = \int^{A:\C} \C(X,A\otimes Y) \times \C (A\otimes
	R,S)
.\] 
\end{theorem}

When $\C$ is small, the existence of sets of coend lenses of each type is guaranteed by the
cocompleteness of $\set$. When $\C$ is not small, and the lens types correspond to coends
indexed by a large category, we must verify that these sets exist by some other means
(by, for example, giving a $\set$ isomorphism).
Fortunately, this is not difficult for the categories of interest in this work.

\begin{definition}[Tensor composition of coend
lenses]\label{lensTensor}
Let $[v,u]:(X,S)\rightarrow (Y,R)$ and $[v',u'] : (X',S')\rightarrow
(Y',R')$ be coend lenses. Define $[v,u]\otimes[v',u'] : (X\otimes
X',S\otimes S')\rightarrow (Y\otimes Y',R\otimes R')$ to be
\begin{center}
\tikzStart\input{tikz/coLensT}\tikzEnd 
\end{center}
Explicitly, $[A,v,u]\otimes [A',v',u']$ is given by
\[
	\Big[ (A \otimes A', \id_A\otimes s_{Y,A'} \otimes \id_{Y'})\circ (v\otimes v'), (u\otimes
	u')\circ (\id_A\otimes s_{A',R} \otimes\id_{R'}) \Big] .
\]
\end{definition}

\begin{theorem}[$\lens_\C$ is symmetric monoidal]
The category $\lens_\C$ is symmetric monoidal with the tensor given
in Definition \ref{lensTensor}, monoidal unit $I = (I_\C,I_\C)$,  and with structural morphisms inherited from $\C$
given by
\begin{align*}
	\alpha_{(X,A),(Y,B),(Z,C)} &= [\alpha_{X,Y,Z}, \alpha^{-1}_{A,B,C}]
\\
	\lambda_{(X,A)} &= [\lambda_X, \lambda_A^{-1}]
\\
	\rho_{(X,A)} &= [\rho_X, \rho_A^{-1}]
\\
	s_{(X,A),(Y,B)} &= [ s_{X,Y}, s_{B,A}]  .
\end{align*}
\end{theorem}

\begin{lemma}
$\lens_\set$ is isomorphic to $\conlens$. (More generally, when $\otimes$ is cartesian, $\lens_\C$ is isomorphic to an appropriately generalised definition of $\conlens_\C$.) \qed
\end{lemma}

\subsection{Towards generalising open games }\label{sec:genGames}

We could, at this point, attempt to define a (generalised) open game $\G :
(X,S)\rightarrow (Y,R)$ over a symmetric monoidal category $\C$ as
\begin{enumerate}
	\item A set $\Sigma$ of strategies;
	\item A play function $\play : \Sigma \rightarrow \lens_\C \big((X,S),(Y,R)\big)$; and
	\item A best-response function $\br : \C(I,X) \times \C(Y,R) \rightarrow
		\rel(\Sigma)$.
\end{enumerate}

Call such generalised open games \textit{interim open games} (for they will not live long).
Sequential composition and tensor composition of interim open games could be defined much as
we did for concrete open games. The problems begin to arise 
when one attempts to prove that this definition results in a symmetric monoidal category. 

In proving that the associator was natural in $\conlens$, we used the fact that the
monoidal tensor in $\set$ is cartesian. If the tensor of $\C$ is \emph{not} cartesian, the
local context of $\G$ in $\G\otimes(\HH\otimes\K)$ is different to the local context of
$\G$ in $(\G\otimes\HH)\otimes\K$. Let
\begin{align*}
	\G &: (X_1,S_1) \rightarrow (Y_1,R_1)
	\\
	\HH &: (X_2,S_2) \rightarrow (Y_2,R_2)
	\\
	\K &: (X_3,S_3) \rightarrow (Y_3,R_3)
\end{align*}
be interim open games, $p\in \C(I,X_1\otimes X_2 \otimes X_3)$, $k \in \C (Y_1\otimes
Y_2\otimes Y_3, R_1\otimes R_2 \otimes R_3)$, and $(\sigma,\tau,\mu)\in
\Sigma_\G\times\Sigma_\HH\times\Sigma_\K$. The local context of $\G$ in
$\G\otimes(\HH\otimes\K)$ is given by
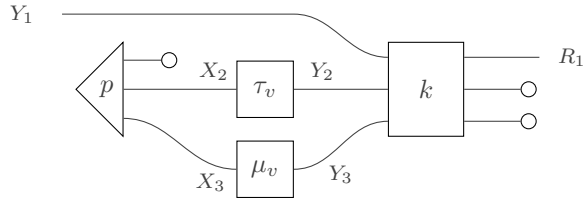
\begin{center}
	\tikzStart\input{tikz/brokenContext1}\tikzEnd
\end{center}
whilst the local context of $\G$ in $(\G\otimes\HH)\otimes\K$ is given by
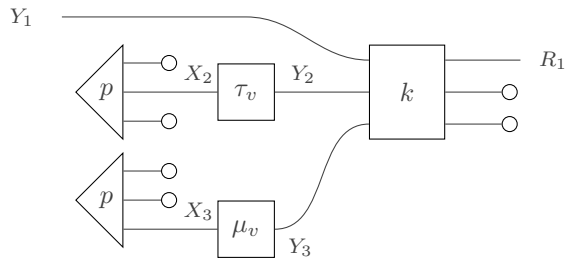
\begin{center}
	\tikzStart\input{tikz/brokenContext2}\tikzEnd 
\end{center}

In general, these morphisms are \emph{not} the same. In the case where $\C$ is the Kleisli category of the distribution monad, the first morphism contains information about correlations
between the types $X_2$ and $X_3$ whilst the second morphism does not. Consequently, the
distinction between these two local contexts for $\G$ is substantive. Fortunately, coend
lenses also provide a solution to this problem.

The high-level approach for defining a category of generalised open games is to use as few
`deleting' maps as possible. We do this by `hiding' information in the bound variable of a
coend lens whenever we would otherwise delete it. A consequence of this approach is that
the correct definition of a `context' for generalised open games is quite abstract, but we
will see that this abstractness allows for more elegant proofs and, in any case, disappears
when dealing with the categories we are actually interested in. 

\subsection{States, continuations, and contexts}\label{sec:genContexts}

In this section we define a generalised notion of \textit{context} for open games. Observe
that a state $[s,s'] \in \lens_\C (I, (X,S))$ has the form
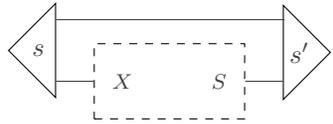
\begin{center}
\tikzStart\node (s) [2state]{$s$};

\path (s.-60) +(0.5,0) node (g)[dbox,anchor=west]{};
\draw (s.-60) -- (g.west) node[label=east:{\fns$X$}]{};

\path (g.east) +(0.5,0) node (s')[2effect,anchor=240]{$s'$};
\draw (s'.240) -- (g.east) node[label=west:{\fns$S$}]{};

\path (s'.west |- s.60) node (m)[inner sep=0]{};
\draw (s.60) -- (m.east);
\tikzEnd 
\end{center}
More verbosely, a state $s\in\lens_\C(I,(X,S))$ is the equivalence class of a choice of
type $A:\C$ together with a state $s:\C(I,A\otimes X)$ in $\C$ and an effect $s': \C(A\otimes
S,I)$ in $\C$. A useful interpretation of states in $\lens_\C$ is as a
\textit{history/cohistory} pair. (Cohistories are not yet well understood. They
make proofs easier, but vanish in categories which make game-theoretic sense.)

An effect $[e,e'] \in \lens_\C ((Y,R),I)$ has the form
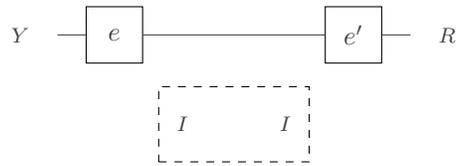
\begin{center}
\tikzStart\input{tikz/lensEffect}\tikzEnd
\end{center}
Concerning effects, we have the following result.
\begin{lemma}\label{lenseffect}
	$\C(Y,R) \cong \lens_\C ((Y,R),I)$
\end{lemma}
\begin{proof}
	The isomorphism $ i : \C(Y,R)\rightarrow \lens_\C((Y,R),I)$ is given by
	\[
		i(f:Y\rightarrow R) = [R,f,\id_R] = [Y,\id_Y,f].
	\]
\end{proof}
\noindent This result captures the idea that `effects in $\lens_\C$ are outcome functions in
$\C$'.

We can now define \textit{(generalised) contexts} which consist of a coend over a state in
$\lens_\C$ (a history/cohistory pair) and an effect in $\lens_\C$ (an outcome function).
Contexts are therefore members of a \textit{double coend}. This double coend turns out to
be a state in the double lens category $\lens_{\lens_\C}$. From a purely technical
standpoint, using double lenses allows for elegant proofs. From a heuristic perspective,
we will see that the extra bound variable the double lens affords us enables us, in the
case $\C = \kl(D)$, to store information about correlations between variables where we
would otherwise have to take marginals.

\begin{definition}[Context functor]
The \textit{context functor} $\context : \lens_\C \times
\lens_\C^{\op} \rightarrow \set$ is given by
\begin{align*}
\context (\Phi,\Psi) &= \int^{\Theta : \lens_\C} \lens_\C
(I,\Theta\otimes\Phi) \times \lens_\C(\Theta\otimes\Psi, I)
\\
&= \lens_{\lens_\C}(I, (\Phi,\Psi))
\end{align*}
Elements of $\context (\Phi,\Psi)$ are called \textit{contexts}.
\end{definition}
(We use the letters $\Phi, \Psi$ to refer to objects of $\lens_\C$, which are pairs of objects of $\C$.)

As a context $[p,k]\in\context( \Phi,\Psi)$ is just a state in
$\lens_{\lens_\C}$, so it admits a graphical representation as
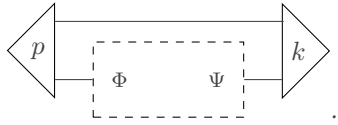
\begin{center}
\tikzStart\node (p)[2state]{$p$};

\path (p.-60) +(0.5,0) node (g)[dbox,anchor=west]{};
\draw (p.-60) -- (g.west) node[label=east:{\fns$\Phi$}]{};

\path (g.east) +(0.5,0) node (k)[2effect,anchor=240]{$k$};
\draw (k.240) -- (g.east) node[label=west:{\fns$\Psi$}]{};

\path (k.west |- p.60) node (mid)[inner sep=0]{};
\draw (p.60) -- (mid.east);
\tikzEnd
.\end{center}
This is neat, and means many of the results in the rest of this section can be carried out
graphically.

\subsection{General open games}

We have now arrived at a level of generality where we can define generalised open games in
a way that is obviously analogous to concrete open games. Given $\Phi,\Psi\in \lens_\C$, an
open game consists of a set of strategy profiles, a family of lenses indexed by the set of
strategy profiles, and a best-response function which takes a context as input and returns a
relation on strategy profiles.

\begin{definition}[Open game]
Let $\Phi,\Psi \in \lens_\C$. An \textit{open game}
$\G:\Phi\rightarrow\Psi$ consists of
\begin{enumerate}
\item A set of \textit{strategy profiles} $\Sigma$;
\item A \textit{play function} $\play:\Sigma\rightarrow \lens_\C
(\Phi,\Psi)$; and
\item A \textit{best-response function} $\br :
\context(\Phi,\Psi)\rightarrow \rel (\Sigma)$.
\end{enumerate}
\end{definition}
The rationale here is much the same as it is with concrete open games. The play function
takes a strategy profile as input and returns a lens describing an open play of the game. Best response takes a context as argument that provides the
information necessary for the game to make informed strategic decisions, and returns a
relation on strategies.

As with concrete open games, we define a notion of \textit{atomic open 
game}:

\begin{definition}
	An \textit{atomic open game} $a: \Phi\rightarrow\Psi$ is an open game such that
	\begin{enumerate}
		\item $\Sigma_a \subseteq \lens_\C (\Phi,\Psi)$;
		\item For all $l \in \Sigma_a$, $\play_a (l) = l$; and
		\item For all contexts $c \in \context(\Phi,\Psi)$, $\br_a(c)$ is
			constant.
	\end{enumerate}
\end{definition}
Atomic open games are uniquely specified by a subset $\Sigma\subseteq\lens_\C(\Phi,\Psi)$
and a selection function $\br : \context(\Phi,\Psi) \rightarrow \pow(\Sigma)$, and we will sometimes specify atomic open games via this data. We refer
to atomic open games simply as \textit{atoms}.

\begin{example}\label{lensid}
The \textit{identity atom} $\id_\Phi : \Phi \rightarrow \Phi$ is
given by $\Sigma = \{ \id_\Phi \}$, $\br(c) = \{ \id_\Phi\}$ for all
$c\in\context(\Phi,\Phi)$.
\end{example}

\begin{example}[Computation]
	Let $f:\C(X,Y)$ and $g:\C(R,S)$ be morphisms in $\C$. Define the atom
$\langle f,g\rangle:(X,S)\rightarrow (Y,R)$ by 
\begin{enumerate}
	\item $\Sigma_{\langle f,g\rangle} = \{ [f,g]\}$; and
	\item $\br_{\langle f,g\rangle}(c) = \{ [f,g] \}$ for all
		$c\in\context\big((X,S),(Y,R)\big)$.
\end{enumerate}
\end{example}

\subsection{Composing open games}

The heuristic for sequential composition of general open games is much the same as for
concrete open games in Subsection \ref{subsec:sComp}. The only difference is that we are now using
coend lenses rather than concrete lenses, and contexts also are slightly different. Best
response of a sequential composite $\HH\circ\G$ is still defined by forming local contexts
for $\G$ and $\HH$.

\begin{definition}[Sequential composition of open games]
Let $\G : \Phi \rightarrow \Psi$ and $\HH : \Psi \rightarrow \Xi$ be
open games. Define $\HH \circ \G : \Phi \rightarrow \Xi$ by
\begin{enumerate}
\item $\Sigma_{\HH\circ\G} = \Sigma_\G \times \Sigma_\HH$,
\item $\play_{\HH\circ\G}(\sigma,\tau) =
	\play_\HH(\tau)\circ\play_\G(\sigma)$,
\item $\br_{\HH\circ\G}([p,k])(\sigma,\tau) =
	\br_\G([p,k\circ\play_\HH(\tau)])(\sigma) \times \br_\HH([\play_\G(\sigma)\circ
	p,k])(\tau)$.
\end{enumerate}
\end{definition}

Given a context $[p,k]\in\context(\Phi,\Xi)$ represented by the diagram
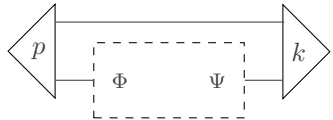
\begin{center}
	\tikzStart\node (p)[2state]{$p$};

\path (p.-60) +(0.5,0) node (g)[dbox,anchor=west]{};
\draw (p.-60) -- (g.west) node[label=east:{\fns$\Phi$}]{};

\path (g.east) +(0.5,0) node (k)[2effect,anchor=240]{$k$};
\draw (k.240) -- (g.east) node[label=west:{\fns$\Psi$}]{};

\path (k.west |- p.60) node (mid)[inner sep=0]{};
\draw (p.60) -- (mid.east);
\tikzEnd
\end{center}
the local context for $\G$ given a strategy $\tau\in\Sigma_\HH$ is given by
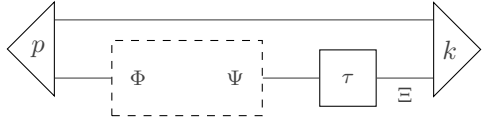
\begin{center}
	\tikzStart\node (p) [2state]{$p$};

\draw (p.-60) -- +(0.75,0) node (g)[dbox,anchor=west]{} node[label=east:{\fns$\Phi$}]{};
\path (g.east) +(0.75,0) node (t)[1morph,anchor=west]{\fns$\tau$};
\draw (t.west) -- (g.east) node[label=west:{\fns$\Psi$}]{};

\draw (t.east) -- node[below]{\fns$\Xi$} +(0.75,0) node (k)[2effect,anchor=240]{$k$};
\draw (p.60) -- (k.west |- p.60);
\tikzEnd
\end{center}
and given a strategy $\sigma\in\Sigma_\G$ the local context for $\HH$ is given by
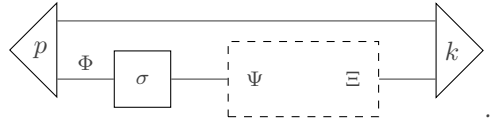
\begin{center}
	\tikzStart\node (p)[2state]{$p$};

\draw (p.-60) -- node[above]{\fns$\Phi$} +(0.75,0) 
	node (s)[1morph,anchor=west]{\fns$\sigma$};
\draw (s.east) -- +(0.75,0) node (g)[dbox,anchor=west]{} node[label=east:{\fns$\Psi$}]{};

\path (g.east) +(0.75,0) node (k)[2effect,anchor=240]{$k$};
\draw (k.240) -- (g.east) node[label=west:{\fns$\Xi$}]{};
\draw (p.60) -- (k.west |- p.60);

\tikzEnd 
\end{center}
In this representation the process of taking a local context is non-arbitrary, and obviously
associative.

\subsection{The tensor of open games}

Again, the heuristic for defining the tensor of open games is much as it was for concrete
open games. We will first formalise the notion of `local context' for tensored general
open games. 

\begin{definition}[Local contexts for tensor composition]
Define the \textit{left local context function}
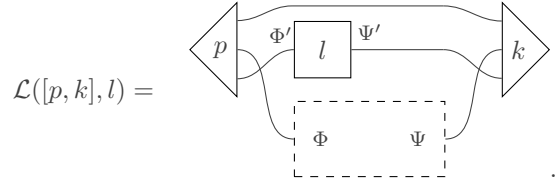
\[
	\leftc_{\Phi,\Phi',\Psi,\Psi'} : \context(\Phi\otimes\Phi',\Psi\otimes\Psi')
\times \lens_\C (\Phi',\Psi') \rightarrow \context(\Phi,\Psi)
\]
by
\begin{center}
\rb{30pt}{$\leftc ([p,k],l) = \quad$}
\tikzStart\input{tikz/leftLocal}\tikzEnd 
\end{center}
Define the \textit{right local context function}
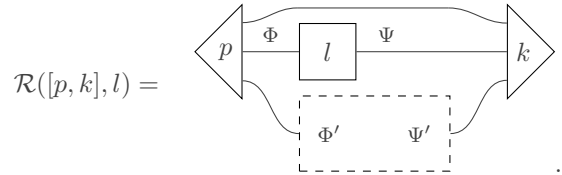
\[
	\rightc_{\Phi,\Phi',\Psi,\Psi'} : \context(\Phi\otimes\Phi',\Psi\otimes\Psi') \times
\lens_\C (\Phi,\Psi) \rightarrow
\context ( \Phi',\Psi')
\]
by
\begin{center}
\rb{30pt}{$\rightc ([p,k],l) = \quad$}
\tikzStart\input{tikz/rightLocal}\tikzEnd 
\end{center}
\end{definition}
We will usually suppress the subscripts of $\leftc$ and $\rightc$ as the types can be
inferred from context.

\begin{definition}[Tensor composition of open games]
Let $\G : \Phi\rightarrow \Psi$ and $\HH : \Phi' \rightarrow \Psi'$ be
open games. Define $\G\otimes\HH : \Phi\otimes\Phi' \rightarrow
\Psi\otimes\Psi'$ by
\begin{itemize}
\item $\Sigma_{\G\otimes\HH} = \Sigma_\G \times \Sigma_\HH$;
\item $\play_{\G\otimes\HH}(\sigma,\tau) = \play_\G(\sigma) \otimes
	\play_\HH(\tau)$ (in $\lens_\C$);
\item Define $\br_{\G\otimes\HH} : \context (\Phi\otimes\Phi',
\Psi\otimes\Psi' ) \rightarrow \rel (\Sigma_{\G\otimes\HH})$ by
\[
\br_{\G\otimes\HH}(c)(\sigma,\tau) = \br_\G(\leftc
(c,\play_\HH(\tau)))(\sigma) \times
\br_\HH(\rightc (c, \play_\G(\sigma)))(\tau)
\]
\end{itemize}
\end{definition}

\subsection{Equivalence of open games}

As in Section \ref{iso}, we need to quotient open games in order to obtain a category.


\begin{definition}[Isomorphism of open games]
	Let $\G,\HH : \Phi\rightarrow\Psi$ be open games. An \textit{isomorphism} of open
	games $\alpha : \G \rightarrow \HH$ is a bijection $\alpha : \Sigma_\G
	\rightarrow \Sigma_\HH$ such that
	\begin{enumerate}
		\item $\play_\G (\sigma) = \play_\HH (\alpha (\sigma))$ for all 
		$\sigma \in \Sigma_\G$; and
		\item For all $\sigma, \sigma' \in \Sigma_\G$ and $c \in \context 
		(\Phi, \Psi)$, $(\sigma, \sigma') \in \br_\G (c)$ iff $(\alpha 
		(\sigma), \alpha (\sigma')) \in \br_\HH (c)$ .
	\end{enumerate}
\end{definition}

\begin{definition}[Equivalence of open games]
	Let $\G,\HH : \Phi\rightarrow \Psi$ be open games. $\G$ and $\HH$ are
	\textit{equivalent}, written $\G\sim\HH$, if there exists an isomorphism $\alpha : \G
	\to \HH$. We write $[\G]$ for the equivalence class of $\G$
	under this relation.
\end{definition}

\begin{lemma}
	Let $\G,\G' : \Phi \rightarrow\Psi$, $\HH,\HH' : \Psi\rightarrow\Xi$, and $\K,\K' :
	\Phi'
	\rightarrow \Psi'$ be open games. Then
	\begin{enumerate}
		\item If $\G\sim\G'$ and $\HH\sim\HH'$, then $\HH\circ\G \sim \HH'\circ\G'$;
			and
		\item If $\G\sim\G'$ and $\K\sim\K'$, then $\G\otimes\K \sim
			\G'\otimes\K'$.
	\end{enumerate}
	\qed
\end{lemma}

Demonstrating equivalence in the cases of interest will always be trivial, and so we
simply specify the witnessing bijection between strategy sets.

\subsection{The category of open games}

That equivalence classes of open games form a category follows easily from the fact that
coend lenses form a category.

\begin{lemma}
	Sequential composition of equivalence classes of open games is associative.
\end{lemma}

\begin{proof}
Suppose we have open games
\begin{center}
\tikzStart
	\node (phi) {$\Phi$};
	\draw[->] (phi.east) -- node[above]{\fns$\G$} +(1,0) node
	(psi)[anchor=west]{$\Psi$};
	\draw[->] (psi.east) -- node[above]{\fns$\HH$} +(1,0) node (xi)[anchor=west]{$\Xi$};
	\draw[->] (xi.east) -- node[above]{\fns$\K$} +(1,0)
	node(up)[anchor=west]{$\Upsilon.$};
\tikzEnd
\end{center}
The equivalence between $(\K\circ\HH)\circ\G$ and $\K\circ(\HH\circ\G)$ will be witnessed by
the isomorphism $\beta : \Sigma_\G \times (\Sigma_\HH \times \Sigma_\K) \to
(\Sigma_\G \times \Sigma_\HH)\times\Sigma_\K$, $(\sigma,
(\tau,\mu)) \mapsto ((\sigma,\tau),\mu)$. Let $\sigma\in\Sigma_\G,\:
\tau\in\Sigma_\HH,$ and $\mu\in\Sigma_\K$. Then
$\play_{(\K\circ\HH)\circ\G}(\sigma,(\tau,\mu)) =
\play_{\K\circ(\HH\circ\G)}((\sigma,\tau),\mu)$ by associativity of composition in
$\lens_\C$. Let $[p,k]\in \context(\Phi,\Upsilon)$ be a context. Then
\begin{align*}
	&((\sigma, (\tau, \mu)), (\sigma', (\tau', \mu'))) \in \br_{(\K \circ \HH) \circ \G} ([p, k]) \\
	\iff\ &(\sigma, \sigma') \in \br_\G \bigg( [p, k \circ \play_\K (\mu) \circ \play_\HH (\tau)] \bigg) \text{ and } (\tau, \tau') \in \br_\HH \bigg( [\play_\G (\sigma) \circ p, k \circ \play_\K (\mu)] \bigg) \\
	&\text{ and } (\mu, \mu') \in \br_\K \bigg( [\play_\HH (\tau) \circ \play_\G (\sigma) \circ p, k] \bigg) \\
	\iff\ &(((\sigma, \tau), \mu), ((\sigma', \tau'), \mu')) \in \br_{\K \circ (\HH \circ \G)} ([p, k])
\end{align*}
\end{proof}

\begin{theorem}
	If $\lens_\C$ exists, there exists a category $\game_\C$ with pairs of 
	objects in $\C$ as objects and equivalence classes of open games as morphisms.
\end{theorem}

\begin{proof}
	All that remains to be checked is that the identity computation defined in
	Example \ref{lensid} is an identity morphism, and this follows from easy checks. 
\end{proof}

\subsection{The symmetric monoidal structure of open games}

We now prove that $\otimes$ is functorial. The proof is a good demonstration of the
utility of coend diagrams. In the commutative squares in the following lemma, the top
path describes how local contexts are formed in, say, $(\HH\otimes\HH')\circ(\G\otimes\G')$
and the bottom path describes how local contexts are formed in
$(\HH\circ\G)\otimes(\HH'\circ\G')$. That the squares commute follows by inspection of the
appropriate coend diagrams.

\begin{lemma}\label{funct}
Suppose we have coend lenses
\begin{center}\tikzStart

\node (tl) {$\Phi$};
\path (tl.east) +(1.5,0) node (tm)[anchor=west]{$\Psi$};
\path (tm.east) +(1.5,0) node (tr)[anchor=west]{$\Xi$};

\path (tl.south) +(0,-0.5) node (bl)[anchor=north]{$\Phi'$};
\path (tm |- bl) node (bm){$\Psi'$};
\path (tr |- bl) node (br){$\Xi'$};

\draw[->] (tl.east) -- node[above]{$l$} (tm.west);
\draw[->] (tm.east) -- node[above]{$m$} (tr.west);

\draw[->] (bl.east) -- node[above]{$l'$} (bm.west);
\draw[->] (bm.east) -- node[above]{$m'$} (br.west);
\tikzEnd\end{center}
The following diagrams commute:
	\begin{enumerate}
		\item \hfill
			\begin{center}\tikzStart\input{tikz/lensF1}\tikzEnd\end{center}

		\item \hfill
			\begin{center}\tikzStart\input{tikz/lensF2}\tikzEnd\end{center}

		\item \hfill
			\begin{center}\tikzStart\input{tikz/lensF3}\tikzEnd\end{center}

		\item \hfill
			\begin{center}\tikzStart\input{tikz/lensF4}\tikzEnd\end{center} 
	\end{enumerate}
\end{lemma}

\begin{proof}
The four squares are given respectively by the following equalities of
coend diagrams:
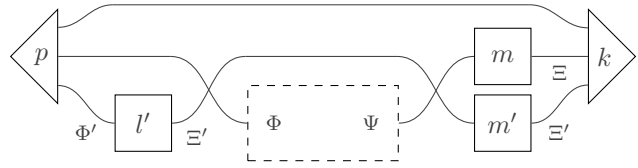
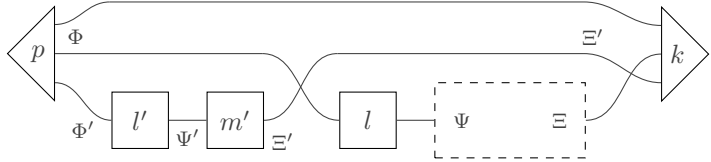
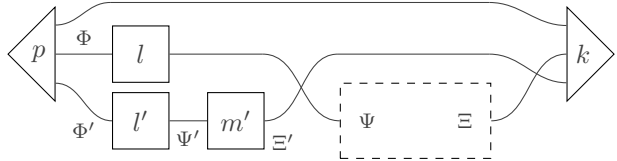
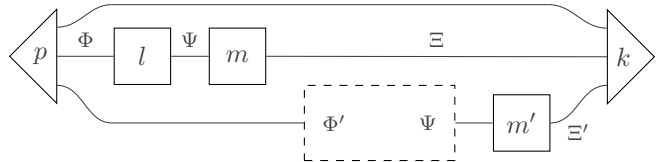
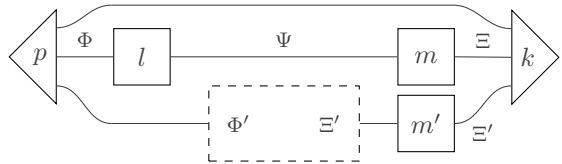
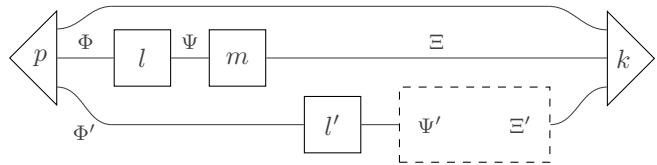
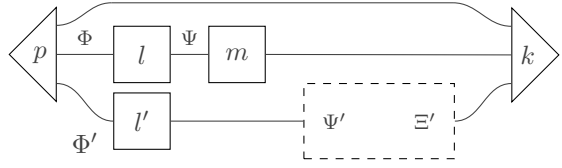
\begin{enumerate}
	\item\hfill
		\begin{center}\tikzStart\input{tikz/sq1}\tikzEnd\vspace{5pt}

$=$\vspace{10pt}

\tikzStart\input{tikz/sq2}\tikzEnd\end{center}

\item\hfill \begin{center}\tikzStart\input{tikz/sq3}\tikzEnd\vspace{5pt}

$=$\vspace{10pt}

\tikzStart\input{tikz/sq4}\tikzEnd\end{center}

\item\hfill
	\begin{center}\tikzStart\input{tikz/sq5}\tikzEnd\vspace{5pt}

$=$\vspace{10pt}

\tikzStart\input{tikz/sq6}\tikzEnd\end{center}

\item\hfill
	\begin{center}\tikzStart\input{tikz/sq7}\tikzEnd\vspace{5pt}

$=$\vspace{10pt}

\tikzStart\input{tikz/sq8}\tikzEnd\end{center}
\end{enumerate}
\end{proof}

Functoriality of the tensor in $\game_\C$ then follows easily.

\begin{corollary}
$\otimes : \game_\C \times \game_\C \rightarrow \game_\C$ is a
functor.
\end{corollary}

\begin{proof}
Suppose we have open games
\begin{center}\tikzStart

\node (tl) {$\Phi$};
\path (tl.east) +(1.5,0) node (tm)[anchor=west]{$\Psi$};
\path (tm.east) +(1.5,0) node (tr)[anchor=west]{$\Xi$};

\path (tl.south) +(0,-0.5) node (bl)[anchor=north]{$\Phi'$};
\path (tm |- bl) node (bm){$\Psi'$};
\path (tr |- bl) node (br){$\Xi'$};

\draw[->] (tl.east) -- node[above]{\fns$\G$} (tm.west);
\draw[->] (tm.east) -- node[above]{\fns$\HH$} (tr.west);

\draw[->] (bl.east) -- node[above]{\fns$\G'$} (bm.west);
\draw[->] (bm.east) -- node[above]{\fns$\HH'$} (br.west);
\tikzEnd\end{center}
Note that $\Sigma_{(\HH\circ\G)\otimes(\HH'\circ\G')} = (\Sigma_\G
\times \Sigma_\HH)\times(\Sigma_{\G'}\times\Sigma_{\HH'})$ and 
$\Sigma_{(\HH\otimes\HH')\circ(\G\otimes\G')} =
(\Sigma_\G \times \Sigma_{\G'}) \times (\Sigma_\HH \times
\Sigma_{\HH'})$. The isomorphism $\beta : (\Sigma_\G \times \Sigma_\HH) \times
(\Sigma_{\G'}\times\Sigma_{\HH'})\to (\Sigma_\G \times
\Sigma_{\G'})\times(\Sigma_\HH \times \Sigma_{\HH'})$ witnessing the equivalence between
$(\HH\circ\G)\otimes(\HH'\circ\G')$ and $(\HH\otimes\HH')\circ(\G\otimes\G')$ is given by
$((\sigma,\tau),(\sigma',\tau'))\mapsto((\sigma,\sigma'),(\tau,\tau'))$.
$\lens_\C$ is symmetric monoidal and, hence,
\[
\play_{(\HH\circ\G)\otimes(\HH'\otimes\G')}((\sigma,\tau),(\sigma',\tau'))=
\play_{(\HH\otimes\HH')\circ(\G\otimes\G)}((\sigma,\sigma'),(\tau,\tau')).
\]
Using Lemma \ref{funct},
\begin{align*}
	&\big(((\sigma, \tau), (\sigma', \tau')), ((\sigma'', \tau''), (\sigma''', \tau'''))\big) \in \br_{(\HH \circ \G) \otimes (\HH' \circ \G')} (c) \\
	\iff\ &(\sigma, \sigma'') \in \br_\G \Big(\context(\Phi,\play_\HH(\tau))\circ \leftc (-,\play_{\HH'}(\tau')\circ\play_{\G'}(\sigma'))(c)\Big) \\
	&\text{ and } (\tau, \tau'') \in \br_\HH \Big(\context(\play_\G(\sigma),\Xi)\circ\leftc(-,\play_{\HH'}(\tau')\circ\play_{\G'}(\sigma'))(c)\Big) \\
	&\text{ and } (\sigma', \sigma''') \in \br_{\G'} \Big( \context(\Phi',\play_{\HH'}(\tau'))\circ \rightc(-,\play_\HH(\tau)\circ\play_\HH(\sigma))(c)\Big) \\
	&\text{ and } (\tau', \tau''') \in \br_{\HH'} \Big( \context(\play_{\G'}(\sigma'), \Xi')\circ \rightc(i,\play_\HH(\tau)\circ\play_\HH(\sigma))(c)\Big) \\
	\iff\ &(\sigma, \sigma'') \in \br_\G \Big(\leftc(-, \play_{\G'}(\sigma'))\circ \context (\Phi\otimes\Phi', \play_{\HH}(\tau)\otimes\play_{\HH'}(\tau')) (c) \Big) \\
	&\text{ and } (\sigma', \sigma''') \in \br_{\G'} \Big(\rightc (-,\play_{\G}(\sigma))\circ \context(\Phi\otimes\Phi', \play_\HH(\tau)\otimes \play_{\HH'}(\tau'))(c) \Big) \\
	&\text{ and } (\tau, \tau'') \in \br_\HH \Big(\leftc( -,\play_{\HH'}(\tau'))\circ \context( \play_\G(\sigma)\otimes \play_{\G'}(\sigma'), \Xi\otimes\Xi') (c) \Big) \\
	&\text{ and } (\tau', \tau''') \in \br_{\HH'} \Big( \rightc(-, \play_\HH(\tau))\circ \context( \play_\G(\sigma)\otimes\play_{\G'}(\sigma'), \Xi\otimes\Xi')(c)\Big) \\
	\iff\ &\big(((\sigma, \sigma'), (\tau, \tau')), ((\sigma'', 
	\sigma'''), (\tau'', \tau'''))\big) \in 
	\br_{(\HH\otimes\HH')\circ(\G\otimes\G')} (c) .
\end{align*}
%
\end{proof}

\begin{definition}
The structural isomorphisms in $\game_\C$ are given by
\begin{align*}
	\alpha_{(X,A),(Y,B),(Z,C)} &= \langle\alpha_{X,Y,Z}, \alpha_{A,B,C}^{-1}\rangle
	\\
	\rho_{(X,A)} &= \langle\rho_X, \rho_A^{-1}\rangle
	\\
	\lambda_{(X,A)} &= \langle\lambda_X, \lambda_A^{-1}\rangle
	\\
	s_{(X,A),(Y,B)} &= \langle s_{X,Y}, s_{B,Y}\rangle  .
\end{align*}
\end{definition}

\begin{lemma}
The structural isomorphisms are natural in $\game_\C$.
\end{lemma}

\begin{proof}
	We show that the associator is natural. Naturality of the other stuctural maps follow by similar
	arguments. Let $\G_i : \Phi_i \rightarrow \Psi_i$ for $i\in\{1,2,3\}$. Note that
	$\Sigma_{\alpha\circ (\G_1\otimes(\G_2\otimes\G_3))} =\big(
		\Sigma_{\G_1} \times
	(\Sigma_{\G_2}\times\Sigma_{\G_3})\big)\times\{\alpha\}$ and
		$\Sigma_{((\G_1\otimes\G_2)\otimes\G_3)\circ\alpha} =
		\{\alpha\}\times\big((\Sigma_{\G_1}\times\Sigma_{\G_2}) \times
			\Sigma_{\G_3}\big)$. The equivalence between $\alpha \circ \big(
			\G_1 \otimes (\G_2\otimes\G_3)\big)$ and $\big(
		(\G_1\otimes\G_2)\otimes\G_3\big) \circ \alpha$ will be witnessed by the
		isomorphism $\big((\sigma, (\tau, \mu)), \alpha\big)
		\mapsto \big( \alpha, ((\sigma,\tau),\mu)\big)$. Let
		$\sigma\in\Sigma_{\G_1}, \tau\in\Sigma_{\G_2}, \mu\in\Sigma_{\G_3},$  and
		$[p,k]\in \context\big((\Phi_1\otimes (\Phi_2\otimes\Phi_3)),
		((\Psi_1\otimes\Psi_2)\otimes\Psi_3)\big)$. We note
		that the local context for $\G_1$ given this data is the same for both
		games. The local context of $\G_1$  is given by
	\begin{center}\tikzStart
	\node (p)[bigstate]{$p$};
\path (p.50) +(1,0) node (tau)[1morph,anchor=west]{$\tau$};
\draw (p.-50) to [out=0,in=180]  node[above,pos=0.8,yshift=2pt]{\fns$\Phi_2$} (tau.west);
\path (tau.east) -- +(1,0) node (con1){};
\path (p.-50 -| con1) node (mu)[1morph,anchor=west]{$\mu$};

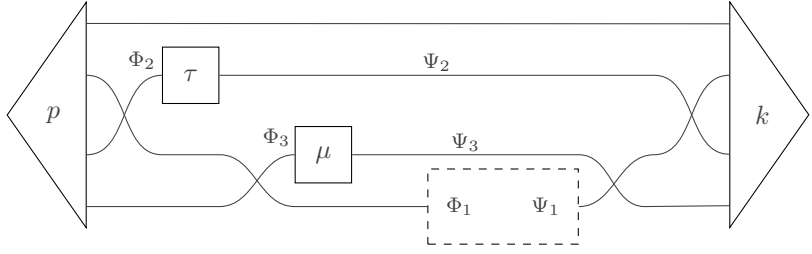
\draw (p.50) to [out=0,in=180] (tau.west |- p.-50)
	-- (tau.east |- p.-50)
	to [out=0,in=180] (mu.west |- p.-70)
	-- (mu.east |- p.-70)
	-- +(1,0) node (g)[dbox,anchor=west]{};

\path (g.west) node [label=east:{\fns$\Phi_1$}]{};
\path (g.east) node [label=west:{\fns$\Psi_1$}]{};

\draw (p.-70) -- (tau.east |- p.-70)
	to [out=0,in=180]  node[above,pos=0.8,yshift=2pt]{\fns$\Phi_3$} (mu.west);

\path (g.east) -- +(1,0) node (con2){};
\path (g.east) -- +(2,0) node (con3){};
\draw (g.east)
	to [out=0,in=180] (con2 |- mu.east)
	to [out=0,in=180] (con3 |- tau.east) node (k)[bigeffect,anchor=130] {$k$};

\draw (p.70) -- (k.110);

\draw (tau.east)
	-- node[above,pos=0.5,yshift=-2pt]{\fns$\Psi_2$} (con2 |- tau.east)
	to [out=0,in=180] (k.230);

\draw (mu.east)
	-- node[above,pos=0.5,yshift=-2pt]{\fns$\Psi_3$} (g.east |- mu.east)
	to [out=0,in=180] (con2)
	; 
	\draw (7.85, -1.217) -- (k.250);

	\tikzEnd\end{center}
	in $\alpha \circ \big(\G_1 \otimes (\G_2\otimes\G_3)\big)$ and by
	\begin{center}\tikzStart
	\input{tikz/ogameass2}
	\tikzEnd\end{center}
	in $\big((\G_1\otimes\G_2)\otimes\G_3\big) \circ \alpha$. This two morphisms are
	evidently equal. Similar diagrams demonstrate that the local contexts for $\G_2$
	and $\G_3$ are the same in both games also. 
\end{proof}

\begin{theorem}
$\game_\C$ is symmetric monoidal.
\end{theorem}

\begin{proof}
	All that remains to be shown is that the Mac Lane pentagon and triangle axioms are
	satisfied, but this follows easily as the underlying category $\C$ is symmetric
	monoidal.
\end{proof}

\subsection{Nice categories of open games}\label{sec:causal}

In this section we show how the notion of `cohistory' collapses when the monoidal unit
$I$ of the underlying monoidal category $\C$ is terminal. With cohistories gone, we will
see that $\game_\C$ has a very natural game-theoretic interpretation.

\begin{lemma}[\cite{mitchell}]\label{terminalunit}
	If the monoidal unit of $\C$ is terminal, then
	$\lens_\C (I,(X,S)) \cong \C(I,X)$.\qed
\end{lemma}

	The isomorphism $i : \C(I,X) \rightarrow \lens_\C (I,(X,S))$ is explicitly given by $ p
	\mapsto [p, !_s]$. In a diagram,
	\begin{center}
	\tikzStart\node (p) [2state]{$p$};
\draw (p.east) -- +(0.75,0) node[label=east:{\fns$X$}]{};
\tikzEnd
	\rb{15pt}{\quad$\mapsto$\quad}
	\tikzStart\node (p)[2state]{$p$};

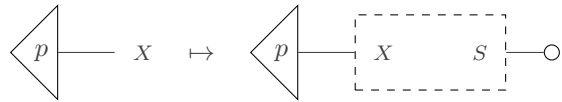
\draw (p.east) -- +(0.75,0) node (g)[dbox,anchor=west]{} node[label=east:{\fns$X$}]{};
\path (g.east) +(0.5,0) node (d)[copy,anchor=west]{};
\draw (d.west) -- (g.east) node[label=west:{\fns$S$}]{};
\tikzEnd
	\end{center}

The following fact appears as \cite[exercise 1.13]{loregian}; thanks to Guillaume Boisseau and Amar Hadzihasanovic for the discussion at \cite{zulip}.

\begin{lemma}\label{adjunctionlemma}
	If $F \dashv U : \D \to \C$ is any adjunction and $G : \D^\op \times \C \to \set$ any functor, then
	\[ \int^{C \in \C} G (F (C), C) \cong \int^{D \in \D} G (D, U (D)) . \]
\end{lemma}

\begin{proof}
	\begin{align*}
		\int^{C \in \C} G (F (C), C) &\cong \iint^{C \in \C, D \in \D} G (D, C) \times \D (F (C), D) &&\text{(ninja Yoneda lemma)} \\
		&\cong \iint^{D \in \D, C \in \C} G (D, C) \times \C (C, U (D)) &&\text{(adjunction, Fubini theorem)} \\
		&\cong \int^{D \in \D} G (D, U (D)) &&\text{(ninja Yoneda lemma)} \qedhere
	\end{align*}
\end{proof}

\begin{lemma}
	If the monoidal unit of $\C$ is terminal, then
	\[
		\context((X,S),(Y,R)) \cong \lens_\C((I,R),(X,Y)) .
	\]
\end{lemma}
	
\begin{proof}
	Let $F : \C \to \lens_\C$ be the embedding $F (X) = (X, I)$. When the monoidal unit of $\C$ is terminal, this functor has a right adjoint $U : \lens_\C \to \C$ that is given on objects by $U (X, S) = X$. On morphisms $U$ is defined by the universal maps
	\[ \int^{A \in \C} \C (X, A \otimes Y) \times \C (A \otimes R, S) \to \C (X, Y) \]
	induced by the dinatural (in $A$) maps $\C (X, A \otimes Y) \times \C (A \otimes R, S) \to \C (X, Y)$, taking $(v, u)$ to $X \overset{v}\longrightarrow A \otimes Y \xrightarrow{!_A \otimes Y} I \otimes Y \cong Y$.
	The adjunction $F \dashv U$ is given by the natural isomorphism
	\[ \int^{A \in \C} \C (X, A \otimes Y) \times \C (A \otimes R, I) 
	\cong \int^{A \in \C} \C (X, A \otimes Y) \times \C (A, I) \cong \C 
	(X, I \otimes Y) \cong \C (X, Y) . \]
	In the previous lemma we take $G : \lens_\C^\op \times \C \to \set$ to be
	\[ G ((\Phi, \Phi'), \Theta) = \C (I, \Theta \otimes X) \times 
	\lens_\C ((\Phi, \Phi') \otimes (Y, R), I) . \]
	Then
	\[ \context ((X, S), (Y, R)) \cong \int^{(\Theta, \Theta') \in \lens_\C} G ((\Theta, \Theta'), \Theta) \cong \int^{\Theta \in \C} G ((\Theta, I), \Theta) \cong \lens_\C ((I, R), (X, Y)) \]
	with the three isomorphisms respectively using Lemmas \ref{terminalunit}, \ref{adjunctionlemma} and \ref{lenseffect}.
\end{proof}


In the case where the monoidal unit of $\C$ is terminal, the type of
best response for an open game $\G :(X,S)\rightarrow(Y,R)$ is equivalently
\[
	\br_\G : \lens_\C ((I,R),(X,Y)) \rightarrow \rel (\Sigma_\G).
\]
We have seen that expressing contexts as states in the double lens category is a good level
of abstraction for categories of open games, allowing for elegant diagrammatic proofs.
From a game-theoretic perspective, however, it will make more sense to express
contexts as equivalence classes $[p,k,\Theta] : \lens_\C ((I,R),(X,Y))$. This is because a
state $p:\C(I,\Theta\otimes X)$ is easily seen to correspond to a \textit{history} for an
open game and the function $k : \Theta\otimes Y\rightarrow R$ acts like an \textit{outcome
function}. In this way, we can specify a context for an open game in much the same way as
we did for concrete open games in \autoref{chap:coGames}.

The coend diagram 
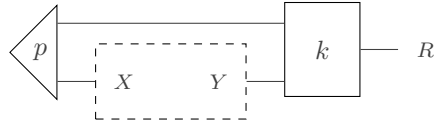
\begin{center}
	\tikzStart\node (p) [2state]{$p$};

\draw (p.-60) -- +(0.5,0) node (g)[anchor=west,dbox]{} node[label=east:{\fns$X$}]{};

\path (g.east) +(0.5,0) node (k)[2morph,anchor=220]{$k$};
\draw (k.220) -- (g.east) node[label=west:{\fns$Y$}]{};

\draw (p.60) -- (k.west |- p.60);

\draw (k.east) -- +(0.5,0) node [label=east:{\fns$R$}]{}; 
\tikzEnd
\end{center}
of a context $[p,k]\in \lens_\C((I,R),(X,Y))$ neatly illustrates that a context is a game
state with a `hole' in it. If we think of a game $\G : (X,S)\rightarrow (Y,R)$ as a player
in a larger game, then $p$ corresponds to the things that have happened in the game before
$\G$ gets to act; $k$ corresponds to what will happen in the game after $\G$ acts; and the
gap in the diagram corresponds to the part of the game where $\G$ gets to influence the
outcome. Alternatively, a context is that which becomes a game once $\G$ has decided which
strategy to play, whereby playing that strategy will fill in the gap in the context.

Given open games $\G : (X,S)\rightarrow (Y,R)$, $\HH : (Y,R)\rightarrow (Z,Q)$, strategies $\sigma\in\Sigma_\G, \tau \in
\Sigma_\HH$, and a
context $[p,k] \in \lens_\C ((I,R),(X,Z))$, the local context for $\G$ in $\HH\circ\G$ is given by
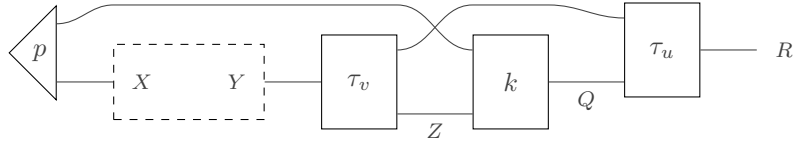
\begin{center}\tikzStart
	\input{tikz/terminalContext1}
\tikzEnd\end{center}
and the local context for $\HH$ is given by
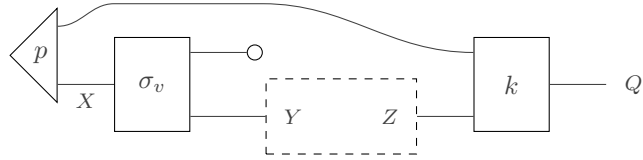
\begin{center}\tikzStart
	\input{tikz/terminalContext2}
\tikzEnd \end{center}
Given another open game $\K : (X',S')\rightarrow(Y',R')$, a context
$[p,k]\in\lens_\C((I,R\otimes R'), (X\otimes X', Y\otimes Y'))$, and a strategy
$\mu\in\Sigma_\HH$, the local contexts for $\G$ and $\HH$ in $\G\otimes\HH$ are given by
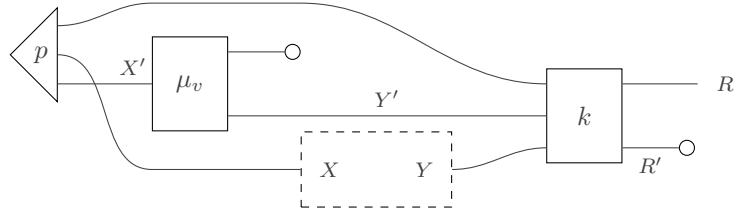
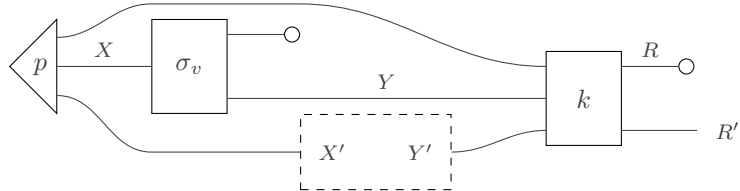
\begin{center}\tikzStart
	\input{tikz/terminalContext3}
\tikzEnd\end{center}
and
\begin{center}\tikzStart
	\input{tikz/terminalContext4}
\tikzEnd\end{center}
respectively.

\section{Bayesian open games}\label{chap:bGames}

In this section we will zero in on open games with a specific lens structure. As
we will show this class of open games will address the shortcomings of
concrete open games. 

\subsection{Commutative monads}\label{sec:bayMonads}

Recall that a monad $T$ over a monoidal category $\C$ is \textit{strong} if it comes with
a \textit{strength} natural transformation $t_{A,B} : A\otimes TB \rightarrow T(A\otimes
B)$ satisfying various coherence conditions.

We have the following result guaranteeing the existence of a large class of coend lens
categories. We refer the reader to~\cite{mitchell} for a much more in-depth discussion of
the following result, and many more examples of when lens categories exist.

\begin{theorem}[\cite{mitchell}]
	If $T$ is a strong monad, then $\lens_{\kl(T)}$
	exists\footnote{In~\cite{mitchell}, lenses over a Kleisli category are called
	\textit{effectful optics}.}.\qed
\end{theorem}

\begin{definition}[Commutative monad]
	Let $T$ be a strong monad with strength $t$ over a monoidal category $\C$. Define
	the \textit{costrength} natural transformation $t'_{A,B} : TA\otimes B\rightarrow
	T(A\otimes B)$ to be the composite
	\begin{center}
		\tikzStart\node (l) {$TA\otimes B$};
\draw[->] (l.east) -- node[above]{\fns$s_{TA,B}$} +(1.5,0)
	node(l2)[anchor=west]{$B\otimes TA$};
\draw[->] (l2.east) -- node[above]{\fns$t_{B,A}$} +(1.5,0)
	node(l3)[anchor=west]{$T(B\otimes A)$};
\draw[->] (l3.east) -- node[above]{\fns$T(s_{B,A})$} +(1.5,0)
	node(l4)[anchor=west]{$T(A\otimes B).$};
\tikzEnd
	\end{center}
	$T$ is \textit{commutative} if the diagram
	\begin{center}
		\tikzStart\input{tikz/comMonad2}\tikzEnd
	\end{center}
	commutes for all objects $A$ and $B$ in $\C$.
\end{definition}

If a monad is commutative then we get that its Kleisli category is symmetric monoidal for
free with the monoidal tensor $\otimes$ (on objects) and unit being the same as in the underlying
category $\C$ .

\begin{lemma}[\cite{power_robinson}]
	If $T$ is a commutative monad over a symmetric monoidal category $\C$, then
	$\kl(T)$ is symmetric monoidal.\qed
\end{lemma}

Commutative monads over $\set$ also come with canonical copy/delete comonoid structures
for every object. Copying $c_X : X\rightarrow T(X\times X)$ is given by 
\begin{center}
	\tikzStart
		\node (l){$X$};
		\draw[->] (l.east) -- node[above]{\fns$\Delta$} +(1.5,0)
			node(m)[anchor=west]{$X\times X$};
		\draw[->] (m.east) -- node[above]{\fns$\eta$} +(1.5,0)
			node(r)[anchor=west]{$T(X\times X)$};
	\tikzEnd
\end{center}
and deleting $d_X : X\rightarrow I$  is given by
\begin{center}
	\tikzStart
		\node (l){$X$};
		\draw[->] (l.east) -- node[above]{\fns$!$} +(1.5,0)
			node(m)[anchor=west]{$\iset$};
		\draw[->] (m.east) -- node[above]{\fns$\eta$} +(1.5,0)
			node(r)[anchor=west]{$T(\iset)$.};
	\tikzEnd 
\end{center}
From this comonoid structure we obtain canonical projections
\begin{center}
	\tikzStart
	\node (l){$X\otimes Y$};
	\draw[->] (l.east) -- node[above]{\fns$\id\otimes d$} +(1.5,0)
		node(m)[anchor=west]{$X\otimes I$};
	\draw[->] (m.east) -- node[above]{\fns$\rho$} +(1.5,0)
		node(r)[anchor=west]{$X$};
	\tikzEnd
\end{center}
and
\begin{center}
	\tikzStart
	\node (l){$X\otimes Y$};
	\draw[->] (l.east) -- node[above]{\fns$d \otimes \id$} +(1.5,0)
		node(m)[anchor=west]{$I\otimes Y$};
	\draw[->] (m.east) -- node[above]{\fns$\lambda$} +(1.5,0)
		node(r)[anchor=west]{$Y$.};
	\tikzEnd 
\end{center}
Crucially, it is \emph{not} guaranteed that the monoidal tensor of $\kl(T)$ is cartesian. 

\subsection{The category of sets and random functions}\label{sec:bayRandom}

We now turn to the category of interest for this section.

The \textit{finitary distribution monad} $D: \set\rightarrow\set$ maps a set $X$ to the
set of finitary probability distributions on $X$ (finitary in the sense that only finitely
many elements are assigned non-zero probability).

\begin{definition}[Finitary distribution monad]
Define $D : \set \rightarrow \set$ by
\[
	D(X) = \Big\{\alpha:X\rightarrow [0,1] \: \Bigm| \: 
	\supp (\alpha)<\aleph_0,\: \sum_{x\in\supp(\alpha)}\alpha(x)=1 \Big\}
\]
where $\supp(\alpha)$ is $\big\{ x\in X \bigm|  \alpha(x)\ne 0 \big\}$, the \textit{support} of
$\alpha$. $D$ acts on morphisms by
\[
	D(f:X\rightarrow Y)(\alpha : D(X))(y) = \sum_{f(x)=y}\alpha (x). 	
\]
The monad structure of $D$ is given as follows. The unit is given by
\[
	\eta_X (x) = \delta_x
\]
where
\[
	\delta_x(x') =
	\begin{cases}
		1	&\text{if } x=x' \\
		0	&\text{otherwise}
	\end{cases}
\]
and the extension $f^\dagger : D(X)\rightarrow D(Y)$ of $f:X\rightarrow D(Y)$ is
\[
	f^\dagger(\alpha) = \sum_{x\in\supp(\alpha)} f(x)(y)  .
\]
\end{definition}

\begin{lemma}
	$D$ is a commutative monad.\qed
\end{lemma}

\begin{corollary}
	$\kl(D)$ is symmetric monoidal with canonical copy/delete comonoids and projection
	maps.
\end{corollary}

The monoidal tensor in $\kl(D)$, which represents joint distributions, is \emph{not} cartesian.

\begin{definition}[Copy/delete comonoid for $\kl(D)$]
	The copying map in $\kl(D)$ is given explicitly by $c_X : X\rightarrow D(X\times
	X)$ where
	\[
		c(x)(x_1,x_2) = \begin{cases}
			1	&\text{if } x=x_1=x_2 \\
			0	&\text{otherwise.}
		\end{cases}
	\]
	The monoidal unit of $\kl(D)$ is terminal, and hence the deleting map $d_X :
	X\rightarrow D(\iset)$ must be given by $d(x)(\star) = 1$. As this map is unique,
	we refer to it as $!$.
\end{definition}

Whenever we are working in $\kl(D)$ we denote $c_X$ and $!_X$ by \begin{center} 
\tikzStart\node (c)[copy]{};

\path (c.east) +(0.5,0.5) node (o1)[label=east:{\fns$X$}]{};
\path (c.east) +(0.5,-0.5) node (o2)[label=east:{\fns$X$}]{};

\draw (c.45) to [out=45,in=180] (o1);
\draw (c.-45) to [out=-45,in=180] (o2);

\path (c.west) +(-0.5,0) node (i1)[label=west:{\fns$X$}]{};
\draw (i1) -- (c.west);
\tikzEnd
 \raisebox{15pt}{\quad and\quad }
  \raisebox{13pt}{\tikzStart\node (d) [copy]{};

\path (d.west) +(-0.5,0) node (i1)[label=west:{\fns$X$}]{};
\draw (i1) -- (d.west);
\tikzEnd}
  \end{center}
respectively.

The canonical projections in $\kl(D)$ correspond to taking marginals.

\begin{definition}[Marginals]
	Let $p : D(X\times Y)$ be a joint distribution. The \textit{left marginal} $p_{X} : D(X)$ is given by $p_X
	(x) = \sum_{y\in\supp( p (x,-))}p(x,y)$. The \textit{right marginal} $p_Y : D(Y)$ is
	given similarly by $p_Y(y) = \sum_{x\in\supp( p(-,y))}p(x,y)$. As diagrams, these
	are given by
	\begin{center}
		\tikzStart\node (p) [2state]{$p$};

\path (p.-60) +(0.5,0) node (d)[copy,anchor=west]{};
\draw (p.-60) -- node[below]{\fns$Y$} (d.west);

\path (d.east |- p.60) node (o1)[label=east:{\fns$X$}]{};
\draw (p.60) -- (o1);
\tikzEnd
		\rb{18pt}{\quad and\quad}
		\rb{3pt}{\tikzStart\node (p)[2state]{$p$};

\path (p.60) +(0.5,0) node (d)[copy,anchor=west]{};
\draw (p.60) -- node[above]{\fns$X$} (d.west);

\path (d.east |- p.-60) node (o1)[label=east:{\fns$Y$}]{};
\draw (p.-60) -- (o1);
\tikzEnd}
	\end{center}
respectively.
\end{definition}

An important operation on probability distributions is \textit{Bayesian updating} where
an agent has some prior distribution (initial belief), makes an observation, and then
updates their prior to a new, posterior distribution.

\begin{definition}[Update operator]
	Let $X$ and $\Theta$ be sets. We think of $\Theta$ as a type which an agent has a
	probabilistic belief about, and $X$ as a type that will be observed by an agent. Define the \textit{update operator}
	$\update_{\Theta} : D(\Theta\times X)\rightarrow (X\rightarrow D(\Theta))$ by
	\[
		\update_{\Theta}(p)(x)(\theta) = \frac
		{p(\theta,x)}{\sum_{(\theta',x)\in\supp (p)}p(\theta',x)}.
	\]
	We also write $p(\theta | x)$ for $\update_\Theta(p)(x)(\theta)$.
\end{definition}

\begin{lemma}\label{naturalUpdate}
	The update operator is natural in $\Theta$. That is, the following diagram
	commutes for any $f : \Theta_1\rightarrow\Theta_2$:
	\begin{center}
	\tikzStart
		\node (tl) {$D(\Theta_1\times X)$};
		\draw[->] (tl.east) -- node[above]{\fns$\update_{\Theta_1}$} +(2,0)
		node (tr)[anchor=west]{$X\rightarrow D(\Theta_1)$};
		\draw[->] (tl.south) -- node[left]{\fns$D(f\times X)$} +(0,-2)
		node (bl)[anchor=north]{$D(\Theta_2\times X)$};
		\draw[->] (bl.east) -- node[below]{\fns$\update_{\Theta_2}$} +(2,0)
		node (br)[anchor=west]{$X\rightarrow D(\Theta_2)$.};
		\draw[->] (tr.south) -- node[right]{\fns$D(f) \circ -$} (br.north);
	\tikzEnd
	\end{center}
\end{lemma}

\begin{proof}
	Let $p \in D(\Theta_1\times X), x\in X, \theta_1\in\Theta_1,$ and
	$\theta_2\in\Theta_2$.
	The top of the square is given by
	\begin{align*}
		\Big( D(f)\circ\update_{\Theta_1}\Big)(p)(x)(\theta_2)
		&= \sum_{f(\theta_1)=\theta_2} \update_{\Theta_1}(p)(x)(\theta_1) \\[2ex]
		&= \frac{\sum_{f(\theta_1)=\theta_2} p(\theta_1,x)}
		{\sum_{\theta_1'\in\supp(p(-,x))}p(\theta_1',x)}. \tag{$\star$}
	\end{align*}
	The bottom of the square is given by
	\begin{align*}
		\update_{\Theta_2}\Big( D(f\times X)(p)\Big) (x)(\theta_2)
		&= \frac{D(f\times X)(p)(\theta_2,x)}
		{\sum_{\theta_2'}D(f\times X)(p)(\theta_2',x)} \\[2ex]
		&= \frac{\sum_{f(\theta_1)=\theta_2}p(\theta_1,x)}
		{\sum_{\theta_2'}\sum_{f(\theta_1')=\theta_2'} p(\theta_1',x)}
		\tag{$\star\star$}.
	\end{align*}
	The result follows, noting that the denominators of $(\star)$ and $(\star\star)$
	are equal.

\end{proof}

\subsection{Bayesian open games}\label{sec:bayOpen}

\begin{definition}[Bayesian open game]
	A \textit{Bayesian open game} is a morphism in $\game_{\kl(D)}$. Explicitly, a
	Bayesian open game $\G : (X,S)\rightarrow (Y,R)$ is given by 
	\begin{enumerate}
		\item A set of strategies $\Sigma$,
		\item A play function
		\[ \play : \Sigma \rightarrow \int^{\Theta : \mathbf{Set}} \big( X\rightarrow
				D(\Theta\times Y)\big) \times \big((\Theta\times R)\rightarrow
			D(S) \big),
		\]
		\item A best-response function
		\[
			\br : \lens_{\kl (D)}((I,R),(X,Y))\rightarrow \rel (\Sigma)
		\]
	\end{enumerate}
	We refer to atoms in the category of Bayesian open games as \textit{Bayesian
	atoms}.
\end{definition}

We unpack the definition of the play function to emphasise that, when we wish to actually
specify a Bayesian open game, it is usually easier to specify $\play(\sigma)$ as the
equivalence class of a pair of morphisms.


\subsection{Bayesian agents}\label{sec:bayAgents}

We will now define \textit{Bayesian agents} which, as with concrete open games, have
constant best-response functions. Bayesian agents capture the notion of rational agents
that
\begin{enumerate}
	\item Have a correct prior about the various types in a game;
	\item Update this prior based on an observation to a new posterior;
	\item Attempt to maximise their expected utility given their posterior.
\end{enumerate}

\begin{definition}[Bayesian agent]
	Let $X,Y$ be sets. The \textit{Bayesian agent} $\A_{(X,Y)} : (X,I)\rightarrow
	(Y,\R)$ is the Bayesian atom given by
	\begin{enumerate}
		\item $\Sigma_\A = X\rightarrow D(Y)$,
		\item$\:$
			
		\begin{center}
			\rb{15pt}{$\play \sigma = [\sigma, !_\R] = $\quad}
			\tikzStart\node (s) [1morph]{$\sigma$};

\path (s.west) +(-0.5,0) node (i1)[label=west:{\fns$X$}]{};
\draw (i1) -- (s.west);

\path (s.east) +(0.5,0) node (g)[dbox,anchor=west]{};
\draw (s.east) -- (g.west) node[label=east:{\fns$Y$}]{};

\path (g.east) +(0.5,0) node (d)[copy,anchor=west]{};
\draw (d.west) -- (g.east) node[label=west:{\fns$\R$}]{};
\tikzEnd
		\end{center}
	\item The selection function $\eps : \lens_{\kl(D)}((I,\R),(X,Y))
		\rightarrow \P(X\rightarrow D(Y))$ is given by
		\begin{align*}
			\eps_\A ([p,k]) =
			\bigg\{ \sigma : X\rightarrow D(Y) \biggm|
				\forall x&\in \supp(p_2),\:\\  \sigma(x)&\in\underset{\alpha\in
				D(Y)}{\arg\max}
		\Big(\exp \Big[ k^\dag 
		(\update_{\Theta}(p)(x),\alpha)\Big]\Big)\bigg\} .
	\end{align*}
	\end{enumerate}
\end{definition}

It is worth explaining this last formula in words. $\A_{(X, Y)}$ represents an agent choosing an element of $Y$ after observing an element of $X$. The context of the decision consists of a set $\Theta$ of unobservable states, a prior joint distribution $p : D (\Theta \times X)$ on unobservable and observable states, and a utility function $k : \Theta \times Y \to D (\R)$ that depends on the unobservable state and the agent's choice. The optimality condition says that for all observations $x$ that the agent could make with nonzero probability, the strategy $\sigma (x)$ maximises the expected value of $k (\theta', -)$, where $\theta' = \mathcal U_\Theta (p) (x)$ is the posterior distribution on the unobservable state given the observation of $x$.

\begin{lemma}
The selection function of a Bayesian agent is well-defined. That is, it is independent of
the choice of representative of the coend equivalence relation.
\end{lemma}

\begin{proof}
	This result follows from the fact that Bayesian updating is natural in the bound
	type of a coend lens (\ref{naturalUpdate}).
\end{proof}

In the next definition we formalise the idea that a player in a game might be assigned a
(game-theoretic) type on which their utility function depends. We can do this simply using
a Bayesian agent and a copying computation.

\begin{definition}
	Let $\A_{(X,Y)} : (X,I)\rightarrow (Y,\R)$ be a Bayesian agent. Define
	$\A_{(X,Y)}^\Delta : (X,I)
	\rightarrow (X\times Y,\R)$ to be the Bayesian open game $(\id_{(X, 1)} \otimes \A_{(X, Y)}) \circ \Delta_X$, or as a string diagram,
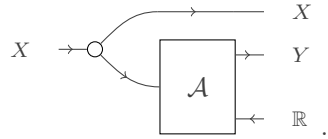
	\begin{center}
		\tikzStart\input{tikz/cAgent.tex}\tikzEnd 
	\end{center}
\end{definition}

\begin{lemma}\label{copyAgent}
	$\A_{(X,Y)}^\Delta$ is explicitly given, up to isomorphism, by
	\begin{enumerate}
		\item	$\Sigma_{\A_{(X,Y)}^\Delta} = X\rightarrow D(Y)$;
		\item $\play_{\A_{(X,Y)}^\Delta} :  \Sigma_{\A_{(X,Y)}^\Delta} \rightarrow
			\lens_{\kl(D)}\big( (X,I),(X\times Y,\R)\big)$ is given by
			\begin{center}
				\rb{0.6cm}{$\play_{\A_{(X,Y)}^\Delta}(\sigma)\quad = \quad$}
				\tikzStart\input{tikz/copyAgent.tex}\tikzEnd
				\rb{0.6cm}{\quad;}
			\end{center}
		\item Let $[p,k] \in \lens_{\kl(D)}\big((I,\R), (X, X\times Y)\big)$. Best
			response is given, up to isomorphism, by
			$\br_{\A_{(X,Y)}^\Delta}([p,k]) =
			\br_{\A_{(X,Y)}}([p,k]')$ where $[p,k]'$ is the context given by
			\begin{center}
			\tikzStart\input{tikz/copyContext.tex}\tikzEnd 
			\end{center}
	\end{enumerate}
\end{lemma}

\begin{proof}
	This result follows from definition chasing, noting that $\A_{(X,Y)}$ is the only
	component with non-trivial strategy profile set.
\end{proof}

\begin{lemma}\label{tensor-agents-lemma}
	Let $\A_{(X_i,Y_i)}$ be Bayesian agents for $i\in\{ 1,\cdots, n\}$. Then
	$\A^{\Delta^n} =\bigotimes_{i=1}^n \A_{(X_i,Y_i)}^\Delta$ is explicitly given as follows.
\begin{enumerate}
	\item $\Sigma_{\A^{\Delta^n}} = \prod_{i=1}^n (X_i\rightarrow D(Y_i))$;
	\item The play function 
		\[
			\play_{\A^{\Delta^n}}: \Sigma_{\A^{\Delta^n}} \rightarrow
			\lens_{\kl(D)}\bigg( \big( \prod_{i=1}^n X_i, I \big),
			\big( \prod_{i=1}^n (X_i\times Y_i), \R^n \big)\bigg)
		\]
		is given by
		\begin{center}
			\rb{55pt}{$\play_{\A^n_\circ}(\sigma_1,\cdots,\sigma_n) \quad=\quad$}
			\tikzStart\input{tikz/cGameTensor}\tikzEnd
		\end{center}
	\item Let $[p,k]\in \lens_{\kl(D)}\Big( (I, \R^n), (\prod_i X_i, \prod_i
		X_i\times Y_i)\Big)$. Best response
		\[
		\br_{\A^{\Delta^n}}([p,k])(\sigma_1,\cdots,\sigma_n)
		\]
		is given, up to isomorphism, by
		\begin{align*}
		\prod_{i=1}^n\bigg\{
		\sigma_i : X_i\rightarrow D(Y_i) \biggm|
		\forall x_i &\in \supp(p^{\sigma_{-i}}), \\
		\sigma_i(x_i) &\in \underset{\alpha_i\in D(Y_i)}{\arg\max}
		\Big( \exp \big[ k^{\sigma_{-i}} \big( \update_{\Theta_{-i}}
		(p^{\sigma_{-i}})(x_i), \alpha_i \big] \Big) \bigg\}
		\end{align*}
\end{enumerate}
where $[p^{\sigma_{-i}}, k^{\sigma_{-i}}]$ is the context  given by
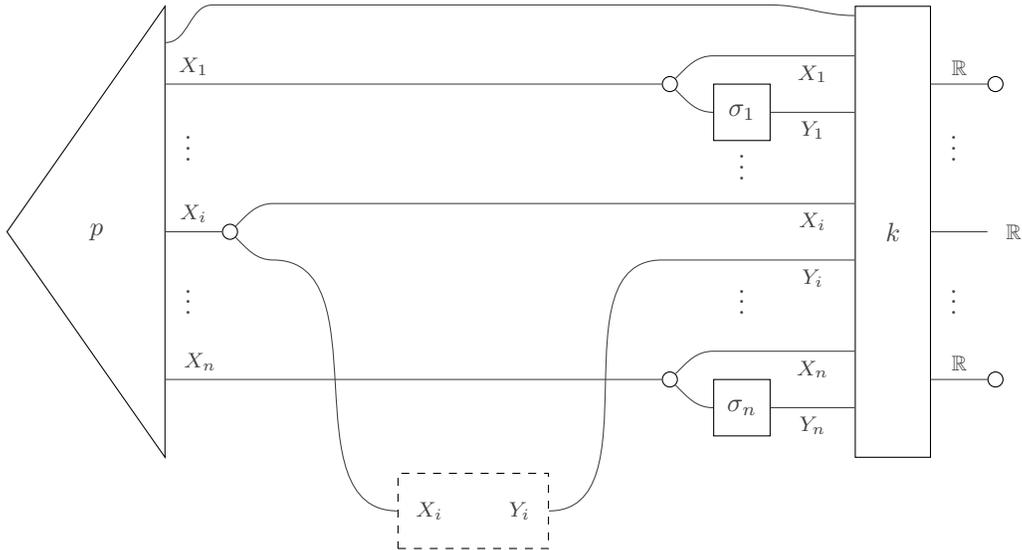
\begin{center}
	\tikzStart\input{tikz/cGameTensor2}\tikzEnd 
\end{center}
\end{lemma}

\begin{proof}
	
	(1) and (2) follow easily from definitions. As for (3), we need to prove that the local
	context for each $\A_{(X_i,Y_i)}$ is $[p^{\sigma_{-i}},k^{\sigma_{-i}}]$. Note
	that the previous Lemma \ref{copyAgent} serves as the base case ($n=1$) for an
	induction argument. The result then follows easily by considering that
	\[
		\br_{\A^{\Delta^n}}([p,k])(\sigma_1,\cdots,\sigma_n) = 
		\br_{\A^{\Delta}_{(X_1,Y_1)}}([p,k]_1)(\sigma_1) \times
		\br_{\bigotimes_{i=2}^n \A^\Delta_{(X_i,Y_i)}}
		([p,k]_{-1})(\sigma_{-1})
	\]
	and applying the inductive hypothesis, where $[p,k]_{-1}$ is the context
	\begin{center}\tikzStart
	\input{tikz/almostDone.tex}
	\tikzEnd\end{center}
\end{proof}

\subsection{Decisions under risk}
In this section we introduce another type of situation involving a Bayesian agent that can
be modelled using Bayesian open games.

A \textit{decision problem under risk} is a decision problem for which one can sensibly
assign probabilities to possible outcomes. A good example is roulette. When making a bet
in roulette, you can calculate the likelihood of success and also your expected return on
any bet. Decision problems under risk are generally represented by Bayesian open games
constructed from computations and precisely one Bayesian agent. A simple subclass of
decision problems under risk are represented by Bayesian open games of the form

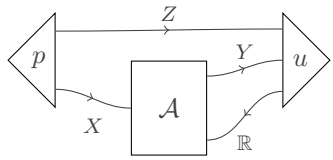
\begin{center}
	\tikzStart\input{tikz/uDecision}\tikzEnd
\end{center}
in which an agent $\A$ attempts to maximise their outcome which is, in part, dependent on
the type $Z$ which $\A$ does not observe.

We now give a fully worked out example of a Bayesian open game in which an agent has a
prior, makes an observation, updates their prior as a consequence of that observation, and
then makes a prediction based on their posterior.

\begin{example}[Biased coin]\label{Exp:BiasedCoin}
	Suppose we give an agent $\A$ a biased coin which lands on one side 75\% of the
	time and the other side 25\% of the time. It is not known which side the coin is
	biased towards, but it is known that it is equally likely to be biased towards
	heads as towards tails. $\A$ flips the coin whilst another identical 
	coin (i.e.~another coin biased the same way) is flipped in secret. $\A$ observes her coin
	flip and is then asked to predict which side up the secret coin landed. If she is
	correct she receives an outcome of $1$ with probability 1. If she is wrong she
	receives an outcome of $0$ with probability 1. The optimal strategy for $\A$ is to
	guess that the coin flipped in secret will land the same way up as the coin she
	flipped. If, for instance, $\A$'s coin comes up heads, then there is a 75\% chance
	that both coins are biased towards heads. Consequently the coin flipped in secret
	is more likely to show heads. A symmetric argument applies if $\A$'s coin shows
	tails.
	
	We can represent this game using the open game
	\begin{center}
		\tikzStart\input{tikz/cFlip}\tikzEnd
	\end{center}
	where \begin{align*}
		p:D(\{H,T\}^2) = \:\:&\frac{1}{2}\bigg( \frac{9}{16}(T,T) +
		\frac{1}{16}(H,H) + \frac{3}{16}(T,H) + \frac{3}{16}(H,T) \bigg)
		\\
				     &+ \frac{1}{2}\bigg( \frac{9}{16}(H,H) +
					     \frac{1}{16}(T,T) + \frac{3}{16}(H,T) +
				     \frac{3}{16}(T,H) \bigg)
		\\
			=\:\: &\frac{5}{16}(H,H) + \frac{5}{16}(T,T) +
				     \frac{3}{16}(T,H) + \frac{3}{16}(H,T)
	\end{align*}
	and
	\begin{align*}
	u &: \{H,T\}^2 \rightarrow D\R
	\\
	(x,y) &\mapsto \begin{cases}
		\delta_1 &\text{if } x=y \\
		\delta_0 &\text{otherwise.}
	\end{cases}
	\end{align*}
Explicitly, the game is given by $\G := u\circ(\A\otimes\id_{\{H,T\}})\circ p$. Note that
$\Sigma_\G \cong \Sigma_\A = \{H,T \} \rightarrow D(\{H,T\})$. Also note that there is
precisely one context for $\G$ since its type is $I \to I$ (that is, $(1, 1) \to (1, 1)$) and, moreover, as the best-response functions for $u,
id_{\{H,T\}},$ and $p$ are trivial, the best-response function for $\G$ is isomorphic to the
constant relation
\begin{align*}
	\br_\A([p,k]) = \bigg\{ \sigma : \{H,T\} \rightarrow D(\{H,T\}) \biggm| \:\:\forall
		&x\in
		\{H,T\}, \\ \sigma (x) &\in \arg\max_{\alpha\in D(\{H,T\})} \exp[k(
	\update_{\{H,T\}}(p)(x),\alpha)]\bigg\}.
	\end{align*}

The posterior $\update_{\{H,T\}}(p)(H)(H)$ is given by
\begin{align*}
	\update_{\{H,T\}}(p)(H)(H) &=
	\frac{p(H,H)}{\sum_{(\theta,H)\in\supp(p)}p(\theta,H)}
	\\
				   &= \frac{5}{8} 
\end{align*}
and hence $\update_{\{H,T\}}(p)(H)(T) = \frac{3}{8}$. Similarly, $\update_{\{H,T\}}(p)(T) =
\frac{5}{8} T + \frac{3}{8} H$. It follows that
\begin{align*}
	{\arg\max}_\alpha \Big( \exp [ k^\dagger (\update_{\{H,T\}}(p)(H),\alpha)]\Big) &=
	\{ \delta_H \}
	\\
	{\arg\max}_\alpha \Big( \exp [ k^\dagger (\update_{\{H,T\}}(p)(T),\alpha)]\Big) &=
	\{ \delta_T \}. 
\end{align*}
Hence $\br_\A([p,k])$ is a singleton set containing the strategy $\sigma$ where $\sigma(H)
=\delta_H$ and $\sigma(T) = \delta_T$, as expected.
\end{example}

\subsection{Relating Bayesian Open Games to Bayesian Games}

The example given in the previous section illustrates two important
improvements over the version of open games as introduced in
\cite{ghani2018compositional}. There, everything is deterministic --  the
environment, the players moves etc. Here, the environment as well as the players'
moves can be probabilistic.

Still, we only consider a single player. We could consider examples with more
players and with probabilistic behaviour induced by nature or
games where probabilistic behaviour is key, such as matching pennies. In \cite{ghani2019compositional} a definition of open games is given that can handle mixed strategies.

Instead, we want to focus on another aspect of Bayesian open games, which we
believe is much less obvious, particularly to readers with no
game theory background: From a game-theoretic perspective, we introduced
a new solution concept. For random environments and probabilistic moves, there
is no real difference to standard Nash equilibria. However, there are important
classes of games for which the new equilibrium notion brings material change.

Reconsider Example \ref{Exp:BiasedCoin}. Here, the agent has some prior
information about the nature of the coin. Given observations from the coin and the prior
information, the agent updates beliefs about the true nature of the coin and
maximises accordingly. Thus, we maintain not only an assumption that the
agent maximises but also how she deals with information.

The example is very simple. After all, there is only one agent. Our interest
lies of course in the interaction of such Bayesian agents. While the strategic
reasoning is more complicated in such situations the overall structure is the
same as in Example \ref{Exp:BiasedCoin}: Each agent has access to some prior
information, some (partial) information is revealed, and the agent makes a
decision. 

In the game-theory literature such a game is called a `Bayesian game' or also
`game in Bayesian form' and the
solution concept we sketched is called `Bayesian Nash Equilibrium'. Next, we
introduce the standard economic notion of such games, then provide a simple
example, an auction, and followed by a general construction that allows us to
translate an arbitrary Bayesian game into a Bayesian open game.  

\subsubsection{Games in Bayesian form}

Denote with $\Gamma^B = \big(N , (A_i)_{i=1}^N, (\Theta_i)_{i=1}^N, \pi ,
(u_i)_{i=1}^N \big)$ a Bayesian game. Before we explain the components we add
some useful notation: we write $\Theta_{-i}=  \prod_{j \in N, j \neq i}
\Theta_j$ for the possible combinations of
$\Theta_j$ other than $i$.

\begin{itemize}
\item $N$ denotes the number of players.
\item $A_i$ defines a set of actions available to the players.\footnote{Note, we
  intenionally refer to this as actions and not as strategies. This will be clear
  once we compare the Bayesian game to a normal-form game.}
\item $\Theta_i$ defines the a set of (player) types each player $i$ can have.
\item $\pi : D \left( \prod_{i = 1}^N \Theta_i \right)$ is a joint prior
  distribution on the types that is common knowledge.
\item $u_i \colon \prod_{i=1}^N \Theta_i \times \prod_{i=1}^N A_i \to \R$
  defines a utility for any profile of player types and any profile of actions.  
\end{itemize}

This definition is rather cryptic at first. So let us try to dissect its components.
Then we will compare it to the definition of a normal form which we
introduced in \ref{normalform}. This hopefully further helps to make
sense of it.

Let us begin with $\Theta_i$, the player type, which is probably most obscure
and most central at the same time. One way to think of types is imagining player
$i$ having different realisations or versions, one of which ends up playing the
game. Each realisation, or type, summarises information relevant for the game.
Often, this information will concern the payoffs.\footnote{The concept of a type
is much more expressive. It can express all sorts of uncertainty with respect to
a game. We will not discuss this here as it would lead us too far astray.
Compare, for instance, Chapters 9-11 in \cite{maschler} for more background on
the notion of type.} For example, player $i$ may
consider to bid in an auction. His type determines his evaluation for the good, that is, how much he values owning the good.
It is easily conceivable that $i$ may have different evaluations before the
auction really takes place.

This example also illustrates another aspect of this structure. While player $i$
may have different types, before he has to choose his action, he typically will
know his type. So, once the bidding starts, he knows how much he values the
good. But crucially, he does not know the types of the other players,
only a probability distribution -- conditional on his own observed type. The
last aspect also already alludes to the role of Bayesian updating: Given his
own observed type, a player may refine his belief about the other players'
types.

The payoff function $u_i$, as with a normal form, maps choices of players into a
payoff. Here, however, player $i$'s type may affect the payoff. Think again
about an auction, naturally the bidding behaviour will affect $i$'s payoff. But
so does his type, the evaluation for the good. Note that the definition above
includes the possibility that the whole realisation of types affects $i$'s
payoff. In the case of bidding for a private good, only $i$'s type will
typically be relevant. 

Lastly, $A_i$ represents the set of actions player $i$ can take in the game.
Note, that we intentionally refer to this as \emph{actions} and not as strategies. To
understand this it is best to compare the above definition to a normal-form
game.

Recall its notation: $\Gamma = \big(N , (S_i)_{i=1}^N, (u_i)_{i=1}^N \big)$. Here,
$S_i$ refers to a strategy which is a complete contingent plan for all possible
occasions where $i$ can make a move. Note that $S_i$ can be a shorthand for some
complicated dynamic structure where $i$ moves several times.

Analogously, $A_i$ in the Bayesian game refers to a complete contingent plan
\emph{once} the game `begins', i.e.~after $i$ has learned his type. This could involve a complicated dynamic structure. Crucially, however, this is not the same as a
strategy for the Bayesian game. Why? A strategy of the Bayesian game must
include a contingent plan also for each realisation of $i$'s type!

Formally, a pure strategy for a game in Bayesian form is a mapping $s_i \colon \Theta_i \to A_i$ for all
$\Theta_i$. A behavioural strategy $\sigma_i$ for player $i$ is a mapping $\sigma_i \colon \Theta_i \to D (A_i)$, i.e
for each type of player $i$ a behavioural strategy assigns a probability distribution over the
available actions.\footnote{In the context of the a game in Bayesian form,
  similar to a game in normal form, it is more common to use \emph{mixed
    strategies} of type $D(S_i) = D (\Theta_i \to A_i)$, i.e.~distributions on pure strategies.
  However, as any game in Bayesian form can be equivalently modelled as a game
  in extensive form, and both, behavioural and mixed strategies lead to the same
  equilibria (under mild assumptions automatically fulfilled in our setup), the
  difference is immaterial. The reason for our choice of using behavioural strategies is that open games share
  properties from both standard formulations and it is easier to connect
  behavioural strategies to our formulation. More on that in Section~\ref{sec:extensive-form}.} 

For each player we can now define the (conditional) expected utility for player
$i$ given behavioural strategy profile $\sigma$ with

\[Eu_i(\sigma \mid t_i) \equiv \sum_{t_{-i} \in \Theta_{-i}} p_\pi (t_{-i} \mid t_i)u_i((t_i,t_{-i}),\sigma)\]

Note, that $p_\pi (\cdot | \cdot)$ updates the prior information $\pi$ for player $i$ given
Bayes' rule.
 Equipped with all that we can finally define a Bayesian Nash equilibrium.

\begin{definition}[Bayesian equilibrium]
 A (behavioural) strategy profile $\sigma^*= (\sigma_1^*, \sigma_2^*, ... , \sigma_N^*)$ is a
 Bayesian Nash equilibrium if for each player $i \in N$, each type $\theta_i \in \Theta_i$, and
 each possible action $a_i \in A_i$ it holds that:
 \[Eu_i(\sigma^* \mid \theta_i) \geq Eu_i (a_i, \sigma_{-i} \mid \theta_i)\]
\end{definition}

\subsubsection{An auction example}

Suppose two agents are bidding for a good that is being sold in a first-price sealed-bid auction. Here `first-price' means that whichever agent bids higher receives the good and pays her/his own bid, with the other agent neither gaining nor losing anything. `Sealed-bid' means neither player can observe the other's bid, meaning the bidding is effectively simultaneous.

Both players have a private valuation for the good, which are drawn from a joint
random distribution $\pi : D (\Theta \times \Theta)$, where $\Theta \subseteq [0, \infty)$ is the
range of possible values. As usual, we assume this prior is common knowledge. One
consequence thereof is that
a player knowing their own valuation can update their beliefs about the other's
valuation. Obviously, whether there is something to learn from one's own
valuation about others' valuations depends on the type of good being
auctioned. For instance, if we are bidding on a construction contract, then the
value that contract has for me will be correlated with your evaluation. If, 
however, we are bidding on a painting by some obscure painter, it is less clear that
I can learn something about others' valuations.

The winning bidder's utility is given by their valuation of the good minus the bid that they must pay. (This can be negative if the bid is higher than the valuation.) We must also choose a way to resolve a tie if the bids are equal: we assume that the winner is determined by a fair coin flip. Thus the expected utilities, for private valuations $\theta_i \in \Theta$ and bids $b_i \in \Theta$, are given by
\[ U (\theta_1, b_1, \theta_2, b_2) = \begin{cases}
	(\theta_1 - b_1, 0) &\text{ if } b_1 > b_2 \\
	(0, \theta_2 - b_2) &\text{ if } b_1 < b_2 \\
	(\frac{\theta_1 - b_1}{2}, \frac{\theta_2 - b_2}{2}) &\text{ if } b_1 
	= b_2 .
\end{cases} \]

In order to formalise this situation as a Bayesian game as defined in the previous section, we take $N = 2$, $A_1 = A_2 = \Theta$, and $\pi$.

The Bayesian open game describing this situation is given by the string diagram
\begin{center}
\ctikzfig{tikzit/auction_string}
\end{center}

As a morphism $\G : I \to I$ in the category of Bayesian open games, this has $\Sigma_\G = (\Theta \to D (\Theta)) \times (\Theta \to D (\Theta))$, the set of behavioural strategy profiles for the auction, and $B_\G (*)$ the best-response relation for behavioural strategy profiles.

\subsubsection{General construction for games in Bayesian form}

Generalising the previous example, we can give a construction for converting any
game in Bayesian form into a Bayesian open game of type $I \to I$ that has the same best-response relation.

Consider a general Bayesian game $(N, (A_i), (\Theta_i), \pi, U)$. For each player $i$, recall the copying agent $\mathcal A^\Delta_{(\Theta_i, A_i)} : (\Theta_i, 1) \to (\Theta_i \times A_i, \R)$. Their tensor product has type
\[ \bigotimes_{i = 1}^N \mathcal A^\Delta_{(\Theta_i, A_i)} : \left( 
\prod_{i = 1}^N \Theta_i, 1\right) \to \left( \prod_{i = 1}^N (\Theta_i 
\times A_i), \R^N \right) . \]

The full game is given by
\[ I \xrightarrow{\left< \pi, 1 \right>} \left( \prod_{i = 1}^N \Theta_i, 
1 \right) \xrightarrow{\displaystyle \bigotimes_{i = 1}^N \mathcal 
A^\Delta_{(\Theta_i, A_i)}} \left( \prod_{i = 1}^N (\Theta_i \times A_i), \R^N \right) \xrightarrow{\left< U, \id_{\R^N} \right>} (\R^N, \R^N) \xrightarrow{\varepsilon_{\R^N}} I . \]

\begin{proposition}
	Let $G = (N, (A_i), (T_i), \pi, U)$ be a game Bayesian form, and let
	\[ c = (1, \pi, U) : \mathbb C \left( \left( \prod_{i = 1}^N T_i, 1 
	\right), \left( \prod_{i = 1}^N (T_i \times A_i), \R \right) \right) . \]
	Then $B_{\bigotimes_{i = 1}^N \mathcal A^\Delta_{(T_i, A_i)}} (c)$ is the best-response relation for behavioural strategy profiles of $G$.
\end{proposition}

The main part of the proof of this result is contained in Lemma \ref{tensor-agents-lemma}, which characterises the tensor product of copying agents.

\subsection{Extensive form}\label{sec:extensive-form}

The normal-form representation as well as the representation of Bayesian Games
really are strategy-centric. They both condense the strategic situation in a way that
makes them most amenable to analysis from the point of view of a specific
equilibrium concept such as Nash/Bayesian Nash equilibrium.

Open games have baked in a notion of equilibrium and thus certainly share this focus. And for that
reason, comparisons to classical game theory in this paper have been limited to
these representations.

In classical game theory, an alternative to the normal-form representation of
a game is its \emph{extensive form}. This representation can also be used for
games in Bayesian form. We will not give a definition here, as
standard definitions are cumbersome and will not add much value. What is
important though is what the extensive form represents: it provides a detailed
account about which player knows what at which point and what actions he has
then available. One can think of the extensive form as a detailed
account of how a game and the players' information unfolds. Thus, it provides
much more information than its corresponding normal form, to which each
extensive form can be reduced.

For instance, one reason to consider the extensive form representation of a game
is that it explicitly
captures temporal information, say, if one player moves before the
other player.\footnote{One other important reason to use the extensive form
is to consider different, typically refined versions of Nash equilibria which take
into account the dynamic dimension of games. But this goes beyond our current
paper.}\label{fn:extensive-form} 

Open games as a representation of a strategic situation have much in common with
its extensive form representation. After all they provide an expressive language
to represent a detailed account of a strategic interaction. In the following, we
will make the relationship between Bayesian open games as presented in this
paper and the extensive form representation of Bayesian games more concrete.

We begin with a concrete example that illustrates how open games deal with
sequential moves of players \emph{in the presence of asymmetric information}.

\subsubsection{The market for lemons}

The following is a simple version of a classic model going back to
\cite{akerlof19701970}.\footnote{Note that the original model is not
  game-theoretic in nature. See \cite{Mas-Colell1995} Chapter 13 for a
  discussion and a richer game-theoretic model than presented here.} There are two players, a seller and a buyer. The seller
wants to sell a used car to the buyer; he offers the buyer a price $P$; the buyer
can either accept or reject the offer. Both players have some evaluation for
the car which will affect their profit/utility.

Their evaluation depends on the quality $Q$ of the car. It is drawn from some
fixed distribution $\pi$. Critically, the seller observes the quality before offering
a price to the buyer. The buyer, however, does not directly observe the quality.

The open game below represents this situation.

\begin{center}
\ctikzfig{tikzit/lemon_string}
\end{center}

As Akerlof~\cite{akerlof19701970} observed, due to the asymmetry in information, a transaction may not
occur even though such a transaction would be efficient (and would take place if
both players were perfectly informed). The reason is simple. For each price the
buyer has to take into consideration what quality of car the seller would be
willing to offer. As each seller will only offer prices equal to or above his
evaluation for the car, a low price indicates low quality. However, a high price
does not necessarily reflect high quality; it could just mean that the seller is
charging a high price for an actual `lemon'. Thus, the buyer, given his prior
will update his estimate of the quality for each price
offered. And depending on the parameters, transactions may only occur for low
quality cars but not for high quality cars. 

Note that, similar to our remark in \ref{subsec:twoPlay}, even though
the interaction (and the representation as an open game) is sequential, the
equilibrium notion is static and does not consider the timing of
moves.\footnote{See remark in Section~\ref{fn:extensive-form}. There are
  refinements of Bayesian Nash Equilibria which take the dynamics into account.
  In the context of the lemon market this is relevant if one considers
  extensions of that model. One prominent example is the role of costly signals
  that the informed side may use in order to reduce the uncertainty of the
  buyer. Cf.~\cite{Mas-Colell1995} Chapter 13.C.}

\subsubsection{A general construction}

To translate an arbitrary extensive form game to a string diagram requires an additional \emph{external choice} operator like the one developed in \cite{morphisms}, which is currently not well understood in the Bayesian setting.
We will instead give a much simpler translation for a certain subclass of extensive form games, roughly those whose trees are perfectly balanced.
Specifically we require:
\begin{itemize}
	\item All paths from the root to a leaf have the same length
	\item All information sets intersect only one level of the tree (where the $i$th level is the set of all nodes that are $i$ steps away from the root)
	\item All information sets on the same level are owned by the same player (the root node is always owned by Nature)
	\item All nodes on the same level have the same number of outgoing edges, labelled by the same edge labels
\end{itemize}
Any extensive form game can be converted into one satisfying these conditions that is equivalent for most game-theoretic purposes, by adding (potentially many) dummy nodes and moves leading to redundant subtrees with only strictly dominated payoffs.
Obviously this is unsatisfactory, and the construction in this section should be thought of as an argument that Bayesian open games have the same `expressive power' as extensive form, and not as a practical translation.

Consider an extensive form game $G$ satisfying these conditions, with $N > 0$ players, and $L > 0$ `ordinary' levels with choice nodes, together with level $0$ (the root) and level $L + 1$ (the leaves).

At an ordinary level $1 \leq i \leq L$, suppose the set of nodes is $X_i$, the information equivalence relation is $\sim_i$ (and so the set of information sets at level $i$ is $X_i / {\sim_i}$), all information sets are owned by player $1 \leq P_i \leq N$, and the set of outgoing edge labels for all nodes is $A_i$.
We summarise the connectivity between level $i$ and level $i + 1$ as a transition function $q_i : X_i \times A_i \to X_{i + 1}$.

The data at the root node is a probability distribution $\pi : D (X_1)$ on the Nature moves.
Finally, $X_{L + 1}$ is the set of leaf nodes, and the assignment of payoffs at leaf nodes is summarised as a utility function $U : X_{L + 1} \to \R^N$.

We translate each level of the tree into a Bayesian open game, which we then sequentially compose together.
An ordinary level $1 \leq i \leq L$ will be translated to an open game of type $\G_i : (X_i, \R^N) \to (X_{i + 1}, \R^N)$, level $0$ to an open game $\G_0 : (1, 1) \to (X_1, \R^N)$ and level $L + 1$ to an open game $\G_{L + 1}: (X_{L + 1}, \R^N) \to (1, 1)$.
Thus, the final result will be an open game $\G = \G_{L + 1} \circ \cdots \circ \G_0 : (1, 1) \to (1, 1)$.

We define $\mathcal G_0$ to be the computation $\left< \pi, ! \right>$, where $\pi : 1 \to D (X_1)$ is the prior considered as a state in the Kleisli category, and $! : \R^N \to D (1) = 1$ is the discard map.
That is to say, it is the atomic open game with $\Sigma_{\mathcal G_0} = \{ [\pi, !] \}$ and $B_{\mathcal G_0} (c) = \{ [\pi, !] \}$ for all contexts $c \in \mathbb C \left( (I, I), (X_1, \mathbb R^N) \right)$.
$\mathcal G_0$ is simply depicted by the string diagram
\ctikzfig{tikzit/extensive_nature_string}
and the optic $[\pi, !]$ by the coend diagram
\ctikzfig{tikzit/extensive_nature_coend}

Similarly, we define $\mathcal G_{L + 1}$ to be the atomic open game with $\Sigma_{\mathcal G_{L + 1}} = \{ U \}$ and $B_{\mathcal G_{L + 1}} (c) = \{ U \}$ for all contexts $c \in \mathbb C \left( (X_{L+ 1}, \R^N), (I, I) \right)$, where $U : (X_{L + 1}, \R^N) \to (I, I)$ is the optic given by $(\R^N, U, \id)$, with the coend diagram
\ctikzfig{tikzit/extensive_payoffs_coend}
$\mathcal G_{L + 1}$ is itself denoted by the string diagram
\ctikzfig{tikzit/extensive_payoffs_string}

The open game $\mathcal G_i$ for $1 \leq i \leq L$ is defined by the string diagram
\ctikzfig{tikzit/extensive_layer_string}
Unwinding this definition, we get that
\begin{itemize}
	\item $\Sigma_{\mathcal G_i} \cong \Sigma_{\mathcal A_{(X_i / {\sim_i}, A_i)}} = X_i / {\sim_i} \to D (A_i)$
	\item For $\sigma_i : X_i / {\sim_i} \to D (A_i)$, $P_{\mathcal G_i} (\sigma_i)$ is the optic $(X_i, \R^n) \to (X_{i + 1}, \R^n)$ given by the coend diagram
	\ctikzfig{tikzit/extensive_layer_coend}
	\item For a context $c = (\Theta, p, k) : (I, \R^N) \to (X_i, X_{i + 1})$, $B_{\mathcal G_i} (c)$ is the constant relation
	\[ \left\{ \sigma_i : X_i / {\sim_i} \to D (A_i) \left|\ \sigma_i ([x_i]_{\sim_i}) \in \arg\max_{a_i \in A_i} \mathbb E [\pi_{P_i} (k (p_1, q (p_2, a_i)))\ |\ p_2 = x_i] \right. \right. \]
	\[ \hspace{-18mm} \left. \phantom{\max_{A_i}} \text{ for all } x_i \in 
	\supp (p_2) \right\} , \]
	where $\mathbb E [\pi_{P_i} (k (p_1, q (p_2, a_i)))\ |\ p_2 = x_i]$ is the conditional expectation of
	\[ I \overset{p}\longrightarrow \Theta \otimes X_i \xrightarrow{\Theta \otimes q (-, a_i)} \Theta \otimes X_{i + 1} \overset{k}\longrightarrow \mathbb R^N \xrightarrow{\pi_{P_i}} \mathbb R \]
	given that the right marginal of $p$ is $x_i$.
\end{itemize}

In order to characterise $\G = \G_{L + 1} \circ \cdots \circ \G_0$ inductively we have a choice to either work forwards from $\G_0$ or backwards from $\G_{L + 1}$, corresponding roughly to forward induction and backward induction in game theory.
We choose the former.
The inductive hypothesis is encapsulated in the following lemma.

\begin{lemma}
	For $1 \leq i \leq L$, consider the extensive form game $G_i$ without specified payoffs given by truncating the original extensive form game $G$ to the first $i+1$ levels ($i$ proper levels plus the root node). Note that the set of leaves of $G_i$ is (labelled by) $X_{i + 1}$. Then the open game $\G_i \circ \cdots \circ \G_0 : (1, 1) \to (X_{i + 1}, \R^N)$ is given by the following:
	\begin{itemize}
		\item $\Sigma_{\G_i \circ \cdots \circ \G_0}$ is the set of behavioural strategy profiles of $G_i$, namely $\Sigma_{\G_i \circ \cdots \circ \G_0} = \prod_{j = 1}^i (X_j / {\sim_j} \to D (A_j))$
		\item $(\G_i \circ \cdots \circ \G_0) (\sigma)$ is the optic $(1, a, !)$, where $a : 1 \to D (X_{i + 1})$ is the probability distribution on the leaf nodes of $G_i$ resulting from playing the behavioural strategy profile $\sigma$
		\item Every context of $\G_i \circ \cdots \circ \G_0$ has the form
		\ctikzfig{tikzit/extensive_ih_context}
		and thus is equivalent to one of the form $(1, *, k)$ where $*$ is the unique distribution on one point and $k : X_i \to D (\R^N)$, by taking $k$ to be the partial application of $k' : \Theta \times X_{i + 1} \to D (\R^N)$ to $p : D (\Theta)$. Then $\br_{\G_i \circ \cdots \circ \G_0} ([*, k]) (\sigma)$ is the set of best responses to $\sigma$ in the extensive form game given by $G_i$ together with payoffs, where the $i$th player's payoff from the leaf node $x \in X_i$ is $\mathbb E [k (x)_i]$ (where $k (x)_i$ is the $i$th marginal of the joint distribution $k (x) : D (\R^N)$).
	\end{itemize}
\end{lemma}

By taking $i = L$ in the previous lemma and then post-composing with $\G_{L + 1}$, we obtain the final result.

\begin{theorem}
	$\G = \G_{L + 1} \circ \cdots \circ \G_0 : (1, 1) \to (1, 1)$ is given by the following:
	\begin{itemize}
		\item The set of strategy profiles of $\G$ is the set of behavioural strategy profiles of the original extensive form game $G$
		\item Every context is equivalent to one of the form $(1, *, *)$, and for a behavioural strategy profile $\sigma$, $\br_\G (1, *, *) (\sigma)$ is the set of best responses to $\sigma$ in $G$
	\end{itemize}
\end{theorem}

\section{Conclusion}

In this last section we will discuss the wider context of this work, as well as some work in progress and future work.

\paragraph{Computer support.} Due to the complexity of the definition of Bayesian open games, in practice computer support is necessary to work with real models in the framework. We have created such a software tool in the form of a Haskell library (available at \url{https://github.com/jules-hedges/open-games-hs}), consisting of a `core' implementation of the monoidal category of Bayesian open games, together with a domain-specific embedded programming language that provides a higher level of description roughly equivalent to string diagrams. We have found this to be a practical tool for modelling, especially for rapid prototyping of models, in a variety of applied domains including auction design, governance modelling and blockchain protocol modelling. The implementation of this tool, and these applied case studies, will be described in a series of future papers.

\paragraph{Non-finitary probability.} In Section \ref{chap:ogames} we gave a
general theory of open games over a monoidal category. And then, in Section
\ref{chap:bGames} specialised it to the Kleisli category of the finite support
probability monad. This restriction to finite support distributions was done for
simplicity, but the general machinery we have introduced is applicable to any
category of probabilistic functions. Examples include the Kleisli categories of
the Giry monads on measurable and Polish spaces \cite{giry82}, the Radon monad
on compact Hausdorff spaces \cite{swirszcz_monadic_functors_convexity}, and the
Kantorovich monad on complete metric spaces
\cite{fritz_perrone_probability_monad_colimit}. In fact, we believe that Bayesian open games can be formulated over any Markov category \cite{fritz_synthetic_approach}, which encompasses all of these examples.

\paragraph{Other solution concepts.} In this paper we have focused on the
solution concept of Bayesian Nash equilibrium. It is one of the central
solution concepts in applied economic modelling as it allows us to model situations of asymmetric information.
It is probably no exaggeration that for most economic situations information is
asymmetric and thus the default. Hence, this paper is an important step in
making the theory of open games practically useful for a wide range of
situations. As a side effect, both ordinary mixed strategy Nash equilibria and
correlated equilibria are obtained as special cases of Bayesian Nash
equilibrium. However, for dynamic games it is common to use equilibrium
refinements such as perfect Bayesian equilibrium or sequential equilibrium, in which individual players' beliefs are represented explicitly.\footnote{See
  Chapter 7 in \cite{maschler} for a general discussion of equilibrium
  refinements and Chapter 9 in \cite{Mas-Colell1995} for a discussion of
  refinements for dynamic games.} It is an open question whether it is possible to extend compositional game theory with these stronger solution concepts.

\paragraph{Behavioural aspects.} A currently-unexplored benefit of Bayesian open
games is that neither maximisation of real numbers nor Bayes' law is involved
in the definition of the categorical structure, and instead both enter when
decisions are defined. This means that the same framework can equally well
accommodate agents who neither perfectly maximise, nor update their beliefs
strictly according to Bayes' law. This could make our framework useful for behavioural game theory, where such alternatives are considered
(see, e.g.~\cite{camerer2011behavioral}).

\subsection{Acknowledgements}

The text of this paper is closely based on the first author's PhD thesis, \cite{bolt_probability_nondeterminism}.
The authors gratefully thank Mitchell Riley, whose suggestion to apply coend lenses to open games was crucial and very timely.

\bibliography{thesisBib}
\bibliographystyle{plainnat}

\end{document}